\documentclass[useAMS,usenatbib]{mn2e}

\usepackage{amsmath, amssymb}
\usepackage{epsfig}         
\usepackage{graphicx, float}
\usepackage{multirow}
\usepackage{tabularx}

\usepackage{natbib}
\usepackage{aas_macros}

\usepackage{url}
\usepackage[pdftex,breaklinks]{hyperref}

\newcolumntype{L}[1]{>{\raggedright\arraybackslash}p{#1}}
\newcolumntype{C}[1]{>{\centering\arraybackslash}p{#1}}
\newcolumntype{R}[1]{>{\raggedleft\arraybackslash}p{#1}}
\usepackage{xcolor,colortbl}
\definecolor{Gray}{gray}{0.85}
\newcolumntype{G}{>{\columncolor{Gray}}r}

\def\Mpc{~{\rm Mpc}}
\def\Mpch{~h^{-1} {\rm Mpc}}

\def\kpc{~\rm kpc}

\def\kms{~\rm{km/s}}

\newcommand{\MI}{\textsc{MS}}
\newcommand{\MII}{\textsc{MS-II}}

\newcommand{\Vector}[1]{\mathbf{#1}}

\newcommand {\lcdm}{$\Lambda$CDM}

\newcommand{\thetaBS}{\theta_{\rmn{BS}\cdot\rmn{S}}}
\newcommand{\thetaHS}{\theta_{\rmn{H}\cdot\rmn{S}}}
\newcommand{\NOA}{N_\rmn{O}{/}N_\rmn{A}}

\newcommand{\gsim}{\raisebox{-0.3ex}{\mbox{$\stackrel{>}{_\sim} \,$}}}
\newcommand{\lsim}{\raisebox{-0.3ex}{\mbox{$\stackrel{<}{_\sim} \,$}}}

\newcommand{\refsec}[1]{Sec.~\ref{#1}}

\newcommand{\reftab}[1]{Table~\ref{#1}}
\newcommand{\reffig}[1]{Fig.~\ref{#1}}
\newcommand{\reffigs}[2]{Figs~\ref{#1}-\ref{#2}}

\newcommand{\figDir}{fig_pdf/}

\newcommand{\ibata}{\hyperlink{labelHypertarget}{Ibata14}}
\newcommand{\ibataTwo}{Ibata14}
\newcommand{\ibataThree}{\hyperlink{labelHypertarget}{Ibata14-b}}

\newcommand{\WangWhite}{\hyperlink{labelHypertarget}{WW12}}

\usepackage[normalem]{ulem}
\usepackage{color}
\definecolor{colorChanges}{rgb}{.0,.3,1.}

\newcommand{\MCn}[1]{#1} 
\newcommand{\MCd}[1]{} 
\newcommand{\MCc}[1]{} 
\newcommand{\MCq}[1]{} 

\newcommand{\MCnn}[1]{} 

\title{A new spin on discs of satellite galaxies}

\author[Cautun et~al.]
{\parbox{\textwidth}{
        Marius Cautun$^{1}$\thanks{E-mail : m.c.cautun@durham.ac.uk},
        Wenting Wang$^{1}$,
        Carlos~S.~Frenk$^{1}$
        and Till Sawala$^{1}$\vspace{.4cm}} \\
$^1$   Department of Physics, Institute for Computational Cosmology, University of Durham, South Road Durham DH1 3LE, UK \\
}

\begin{document}


\maketitle

\begin{abstract}
  We investigate the angular and kinematic distributions of satellite
  galaxies around a large sample of bright isolated primaries in the
  spectroscopic and photometric catalogues of the Sloan Digital Sky
  Survey (SDSS). We detect significant anisotropy in the spatial
  distribution of satellites. To test whether this anisotropy could be
  related to the rotating discs of satellites recently found by Ibata
  et al. in a sample of SDSS galaxies, we repeat and extend their
  analysis. Ibata et al. found an excess of satellites on opposite
  sides of their primaries having anticorrelated radial velocities. We
  find that this excess is sensitive to small changes in the sample
  selection criteria which can greatly reduce its significance. In
  addition, we find no evidence for correspondingly correlated
  velocities for satellites observed on the same side of their
  primaries, which would be expected for rotating discs of
  satellites. We conclude that the detection of 
  \MCnn{rotating planes of satellites in the observational sample of Ibata et al. is not
  robust to changes in the sample selection criteria}. We compare our data to the \lcdm{} Millennium simulations
  populated with galaxies according to the semi-analytic model of Guo
  et al. We find excellent agreement with the spatial
  distribution of satellites in the SDSS data and the lack of a strong signal
  from coherent rotation.
\end{abstract} 




\section{Introduction}
\label{sec:introduction}
It has been known for decades that the 11 ``classical'' satellites of
the Milky Way (MW) define a thin plane \citep{Lynden-Bell1976} and
that some of the fainter satellites, tidal streams and young globular
clusters have an anisotropic distribution reminiscent of this plane
\citep{Metz2009,Pawlowski2012}. Many members of this ``disc of
satellites'' have a common rotation direction and it has been claimed
that the plane is a rotationally stabilized structure
\citep{Metz2008c,Pawlowski2013b}. Similarly, the spatial distribution
of satellites around Andromeda is also thought to be anisotropic
\citep{Koch2006,McConnachie2006}, with 15 out of 27 satellites
observed by the Pan-Andromeda Archaeological
Survey\citep[PAndAS;][]{McConnachie2009} located in a very thin plane
\MCn{in which 13 out of the 15 satellites share the same sense of rotation}
\citep{Ibata2013}.

Anisotropies in the distribution of satellites are a clear prediction
of the $\Lambda$ cold dark matter (\lcdm) paradigm
\citep{Libeskind2005,Zentner2005a,Libeskind2009,Libeskind2011,Deason2011,Wang2013}. Such
flattened satellite distributions, dubbed ``great pancakes'', can
arise from the infall of satellites along the spine of filaments
\citep{Libeskind2005}, which in turn determine the preferential points
at which satellites enter the virial radius of the host halo
\citep{Libeskind2011,Libeskind2014}. Correlated accretion along
filaments has also been ascribed by \citet{Deason2011} as the cause of
the satellite anisotropies observed in the `GIMIC` gasdynamic
simulation \citep{Crain2009}; they found a polar alignment of
satellite discs (with more than 10 bright members) for $20\%$ of the
cases. The flattening effects of anisotropic accretion are greatly
enhanced in the case when subhaloes are accreted in groups
(\citealt{Li2008}; \MCnn{though \citealt{Metz2009b} claimed that this would not explain the MW satellite plane}), although such occurrences are rare for bright satellites
and only become more frequent for less massive subhaloes
\citep{Wang2013}. The imprint of anisotropic accretion is retained in
the dynamics of satellites, with a significant population of subhaloes
co-rotating with the spin of the host halo \citep[][for galactic
haloes; \citealt{Shaw2006,Warnick2006} for cluster mass
haloes]{Libeskind2009,Lovell2011}.
\MCnn{An alternative view that has been put forward is that the satellites do not reside in dark matter substructures, but instead are formed from tidal debris produced during galaxy-galaxy interactions, which could also result in the formation of satellite planes \citep[e.g.][]{Fouquet2012,Hammer2013,Yang2014}.}

Although flattened satellite distributions are common in \lcdm{}, the
degree of flattening of the MW and Andromeda satellites is
atypical. \citet{Wang2013} found that $5-10\%$ of satellite systems
are as flat as the MW's 11 classical satellites but, when the
requirement that the velocities of at least 8 of the 11 satellites
should point within the narrow angle claimed by \citet{Pawlowski2013b}
for the MW satellites, this fraction decreases to ${\sim}
1\%$. \MCnn{\citet{Pawlowski2012b} claimed that there is only a 0.5\% chance that this alignment is due to filamentary satellite accretion.} 
In the case of Andromeda's thin satellite plane,
\citet[][\MCn{but see \citealt{Pawlowski2014c}}]{Bahl2014} found that, while similar spatial distributions of
satellites are quite common in \lcdm{}, there is only a $2\%$ chance
that 13 out of the 15 members in the plane share the same sense of
rotation. In a similar study, \citet{Ibata2014b} found an even lower occurrence rate for 
Andromeda's thin plane in \lcdm{} simulations.

The presence of such highly flattened satellite systems in the Local
Group (LG) raises an important question: are such systems ubiquitous
around other galaxies, or are they a consequence of the large-scale
environment in which the LG is located? \MCn{While being a member of a
  pair rather than an isolated halo seems to make little difference
  for the distribution of satellites \citep{Pawlowski2014d}, the
  effect of the megaparsec-scale environment is still unknown. The
  crucial role of large-scale modes } in determining the properties of
the LG was illustrated by \citet{Forero-Romero2011} who, using
constrained simulations of the local cosmological volume, found that
LG-analogues have highly atypical formation times, assembly histories
and times since last major merger when compared to a sample of similar
mass halo pairs.

Studies of large samples of galaxies are limited to investigating
anisotropies in the satellite distribution with respect to
preferential axes defined by the projected galaxy light. For example,
late-type galaxies have satellite distributions that are close to
isotropic, while the satellites of early-type galaxies are aligned
with the major axis of the galaxy's light
\citep{Brainerd2005,Yang2006,Bailin2008,Agustsson2010,Guo2012,Nierenberg2012}. Such
studies have limited power to constrain the full flattening of the
satellite distribution when such anisotropies are uncorrelated, or only
weakly correlated, with the light distribution of the central host.

Recently, \citet[][hereafter \ibata{}]{Ibata2014} analysed
correlations in the velocites of satellite galaxies observed on
opposite sides of their central host. For a sample selected from SDSS 
(\citealt{Abazajian2009}), they found the most significant effect for
an opening angle of $< 8^\circ$, for which 20 out of 22 satellite
pairs have anticorrelated velocities, suggestive of a rotating disc
that contains ${\sim}50\%$ of the satellite population.
\cite{Ibata2014} reported a significance of $4\sigma$ for a null
hypothesis of an isotropic satellite distribution. They found no such
effect in the Millennium II \lcdm{} simulations.

In this study we compare the angular distribution of satellites around
external galaxies with the predictions of the \citet{Guo2011_SAM}
semi-analytic model of galaxy formation model implemented in the
Millennium (\MI; \citealt{Springel2005}) and Millennium II (\MII;
\citealt{Boylan-Kolchin2009}) simulations. We make use of both
spectroscopic and photometric SDSS data, limiting our analysis to systems of
isolated galaxies with at least one spectroscopic satellite. We use
the axis connecting the position of the brightest spectroscopic
satellite to its host galaxy to measure the angles at which other
satellites appear on the sky. The distribution of this angle is
sensitive to anisotropies in the spatial distribution of satellites,
as we show using a simple disc model. We compare the resulting angular
distribution of satellites with \lcdm{} predictions for different
central host luminosities and find very good agreement between the
two. We also show that the excess of satellite pairs with
anticorrelated velocities found by \ibata{} is not robust to changes
in sample selection and conclude that the known kinematics of
satellites are not in disagreement with \lcdm{} predictions.

The paper is organised as follows: \refsec{sec:data} introduces the
observational and simulation data, as well as the selection criteria
used to identify isolated galaxies and their satellites; in
\refsec{sec:spatial_distribution} we obtain the angular distribution
of SDSS satellites and compare the results with \lcdm{} predictions;
\refsec{sec:velocity_distribution} is devoted to studying kinematical
signatures of satellite discs and on revisiting the excess of
satellite pairs with anticorrelated velocities; we conclude with a
short discussion and summary in \refsec{sec:conclusion}.


\section{Data and sample selection}
\label{sec:data}

We identify isolated galaxies and count their satellites using the
methods described by \citet[][hereafter \WangWhite{}; see also
\citealt{WangW2014}]{WangW2012}. We now briefly introduce these
methods and describe the datasets that we used.

\subsection{SDSS isolated galaxy sample}

We select isolated primary galaxies from the New York University Value
Added Galaxy Catalogue (NYU-VAGC)~1 \citep{Blanton2005}, which is
based on the Seventh Data Release of the SDSS (SDSS/DR7;
\citealt{Abazajian2009}). We require that these galaxies should be
brighter than any companion lying within a projected radius of $r_p =
0.5 \Mpc$ and having line-of-sight velocity difference $c|\Delta z| < 1500
\kms$.  In order to match the selection criteria used by \ibata{}, the
results presented in \refsec{sec:velocity_distribution} use only this
sample of isolated primaries.

For the analysis in \refsec{sec:spatial_distribution} we apply a
further isolation selection criterion that takes into account the fact
that the SDSS spectroscopic sample is incomplete due to fibre-fibre
collisions. To prevent primaries being falsely identified as isolated
because of incompleteness in the spectroscopic catalogue, we search
for further companions using the photometric SDSS catalogue.  We
reject primary candidates if they have a photometric companion which
is not in the spectroscopic catalogue but satisfies the position and
magnitude cuts given above and the probability that its redshift is
equal to or less than the primary is larger than 10\%. For this last
step we use the photometric redshift distributions from
\citet{Cunha2009}.

\subsection{SDSS satellite galaxy sample}
\label{sec:sdss_sats}

For the analysis described in \refsec{sec:spatial_distribution} we
first split the isolated galaxy sample into three subsamples according
to their absolute r-band magnitudes. We use three bins centred on
$M_\rmn{r}=-23$, $-22$ and $-21$, each of width $\Delta M_\rmn{r}=1$,
as shown in \reftab{tab:isolated_galaxy_count}. We count as satellites
all galaxies within a projected radius in the range $20\kpc$ to the
virial radius, $R_v$, with $R_v = 500$, 315 and $150\kpc$, which
correspond to the median virial radii of haloes hosting the galaxies 
found in each of our luminosity bins\footnote{We used the
  \citet{Guo2011_SAM} \MI{} catalogue to find the median virial radii
  of galaxies in each magnitude range. The virial radius of the
  brightest bin is larger than $500\kpc$ but we adopted this value
  since it corresponds to the projected radius used to identify
  isolated primaries.}. Out of all the isolated primaries, we keep
only those which have at least one spectroscopic satellite within a
projected distance between $20\kpc$ and $R_v$ and having a
line-of-sight velocity difference, $c|\Delta z| < 300 \kms$. The
number of isolated galaxies satisfying these criteria is given in
\reftab{tab:isolated_galaxy_count}.

\begin{table}
    \centering
    \caption{ The number of isolated galaxies with at least one
      associated spectroscopic satellite. The \MI{} and \MII{}
      data correspond to the average number of isolated galaxies per
      line-of-sight, because multiple lines-of-sight were used to
      construct the mock data.} 
    \label{tab:isolated_galaxy_count}
    \begin{tabular}{lR{1.2cm}R{1.2cm}R{1.2cm} }
        \hline
        host $M_\rmn{r}$ range  &  SDSS  &  \MI{}  &  \MII{} \\
        \hline
        $-22.5\ge M_\rmn{r} \ge -23.5$  &   4\;211 &  16\;430 &    111   \\
        $-21.5\ge M_\rmn{r} \ge -22.5$  &  16\;532 & 112\;100 &    938  \\
        $-20.5\ge M_\rmn{r} \ge -21.5$  &   8\;519 & 235\;360 & 2\;010  \\
        \hline
    \end{tabular}
\end{table}

We are only interested in isolated primaries with spectroscopically
associated satellites since we want to determine a preferential axis
that can be used to probe anisotropies in the satellite
distribution. The relative position of the satellite with respect to
its host represents such a reference axis, $\Vector{x}_\rmn{BS}$,
since the satellite is more likely to be found along the direction
where there is an excess of satellites. If an isolated galaxy has two
or more spectroscopic satellites associated with it, we choose the
brightest one because the brightest satellites show the largest
degree of anisotropy \citep{Wang2013,Libeskind2014}.

To compute the angular distribution of satellites we use the SDSS/DR8
photometric catalogue \citep{Aihara2011}, which we correct
statistically for background contamination \MCn{using the method
  carefully developed and tested by \WangWhite{} (where further
  details may be found)}. For each isolated galaxy, we identify objects
  brighter than apparent magnitude $r = 21$ that are within a
  projected distance between $20\kpc$ and $R_v$. \MCn{We then use the
    redshift of the primary to convert apparent magnitudes into
    rest-frame $r$ and $g$ magnitudes. Of all potential satellites, we
    only keep those that have rest-frame colors $g-r \le 1$, since
    redder objects are too red to be at the redshift of the primary
    galaxy\footnote{For brevity, we only give here a simplified
      description of the $g-r$ colour cut. The exact cut applied is
      stellar mass dependent and includes an elaborate procedure of
      estimating stellar masses using photometric data. The full
      procedure is described in \WangWhite{}.}
    \citep[][\WangWhite{}]{Lares2011}. It is useful to exclude such
    red galaxies since they add noise without adding signal. This
    colour cut represents a conservative and safe selection, 
    equivalent to a crude cut in photometric redshift. }

  For each of these potential satellites, we calculate the angle,
  $\thetaBS$, with respect to the reference axis,
  $\Vector{x}_\rmn{BS}$, of the system.  We then count the number of
  satellites as a function of the angle $\thetaBS$. This count
  excludes the brightest satellite, for which $\thetaBS=0^\circ$ by
  definition.  \MCn{The background galaxy count is given by the number
    of objects brighter than $r = 21$ having rest-frame colour $g-r
    \le 1$, as evaluated at the redshift of the primary. We estimate
    this background from the survey as a whole. For each bin in
    $\thetaBS$, we subtract the average number of background galaxies
    expected in this area of the sky.}
  \MCn{The background fraction for the three primary samples, from
    brightest to faintest, is $57\%$, $80\%$ and $94\%$.} The excess
  counts with respect to a homogeneous galaxy background are assumed
  to be satellites physically associated with the primary
  galaxy. Finally, results for different primaries are averaged after
  making completeness, volume and edge corrections, as described in
  \WangWhite{}.  \MCn{The measurement uncertainties are estimated
    using 100 bootstrap samples over the primary galaxies.}

The selection of satellites in the sample used in
\refsec{sec:velocity_distribution} is restricted to galaxies with
spectroscopic redshifts following the criteria described in
\ibata{}. For each isolated primary we identify galaxies that are at
least $\Delta M_\rmn{r}^\rmn{Sat-Cen} = 1$~mag. fainter than the
primary and lie within a projected distance between $20\kpc$ and
$R_\rmn{max}=150\kpc$. We further require that the line-of-sight
velocity difference of the satellite be $35\kms \le c|\Delta z| \le
V_0\exp\left(-(R/300\kpc)^{0.8}\right)$, where $V_0=300\kms$ and $R$
is the projected distance from the primary of the satellite
candidate. We also limit the analysis to primary galaxies in the
redshift range $0.002$ to $z_\rmn{max}=0.05$. The final sample
consists of all primaries with two or more satellites satisfying the
above selection criteria. To assess the robustness of the results we
vary each of the selection criteria in turn.

\subsection{Mock \lcdm{} galaxy catalogue}

To construct mock catalogues, we use the semi-analytic galaxy
formation model of \citet{Guo2011_SAM} implemented in the \MI\ 
(\citealt{Springel2005}) and \MII\ (\citealt{Boylan-Kolchin2009}). The
semi-analytic model has been calibrated to reproduce the stellar mass,
luminosity and autocorrelation functions of low redshift galaxies as
inferred from SDSS. The abundance and radial distribution of
satellites predicted by the model is in very good agreement with SDSS
data \citep[\WangWhite{};][]{WangW2014}. The two simulations, the high
resolution \MII{} and the lower resolution but larger volume \MI{},
complement each other well for the purposes of this study. The
\citet{Guo2011_SAM} data are publicly available at
\url{http://www.mpa-garching.mpg.de/millennium}.

We create the simulated catalogues by projecting galaxies along
random sightlines and assigning a redshift according to their
line-of-sight distance and peculiar velocity. We add a Gaussian random
velocity error of $\sigma=15\kms$ to the radial velocity to simulate
the typical SDSS spectroscopic redshift error. We then apply the same
host isolation and satellite identification criteria as in our SDSS
data to obtain a mock sample of isolated primaries and their
satellites. In \refsec{sec:velocity_distribution} we use only satellites
brighter than an absolute magnitude of $M_\rmn{r} = -17$. 

\MCn{To mimic the background in the real data, we only consider as
  background galaxies those with apparent magnitude $r\le21$. The
  background depth is restricted to the size of the simulation cube,
  $100\Mpch$ for \MII{} and $500\Mpch$ for \MI{}. Appendix A4 in
  \WangWhite{} presents extensive tests of the background
  estimation in mocks like ours, and explicitly compares the
  background of a projected simulation cube with that of a full
  light-cone mock. \WangWhite{} found that the only difference
  between the two is the size of the uncertainties, which are larger
  for the light-cone mocks. This reflects the smaller effective volume of light-cone mocks in
  the redshift range of interest compared to the effective volume of projected simulation cubes.}
%

In \refsec{sec:spatial_distribution} we
find the angular distribution of satellites by counting all the
satellites brighter than $r=21$ lying within projected distance
between $20\kpc$ and $R_v$, from which we subtract the average galaxy
background of the mock catalogue. When stacking the counts in each
primary magnitude bin, we assign weights to the primaries so as to
obtain the same redshift distributions in our mock and SDSS
samples. We obtain the same average number of satellites per primary
for both mock and real data. We create multiple mock catalogues using
1000 and 25 random sightlines from the \MII{} and \MI{} respectively.


\section{Spatial distribution of satellites}
\label{sec:spatial_distribution}

In this section we characterise the anisotropies of the satellite
distribution around a large number of primary galaxies. For each
system, we define a reference axis, $\Vector{x}_\rmn{BS}$, given by
the relative position of the brightest satellite with respect to the
primary galaxy, as described in \refsec{sec:sdss_sats}. This reference
axis points towards the direction where an excess of satellites is
expected on average, if such an excess exists. We first test this
approach using a simplified disc model, and then we apply the
method to both observations and mock catalogues.

\subsection{A simplified disc model} 
\label{sec:spatial_distribution:disc_model}

\begin{figure}
     \centering
     \includegraphics[width=\linewidth,angle=0]{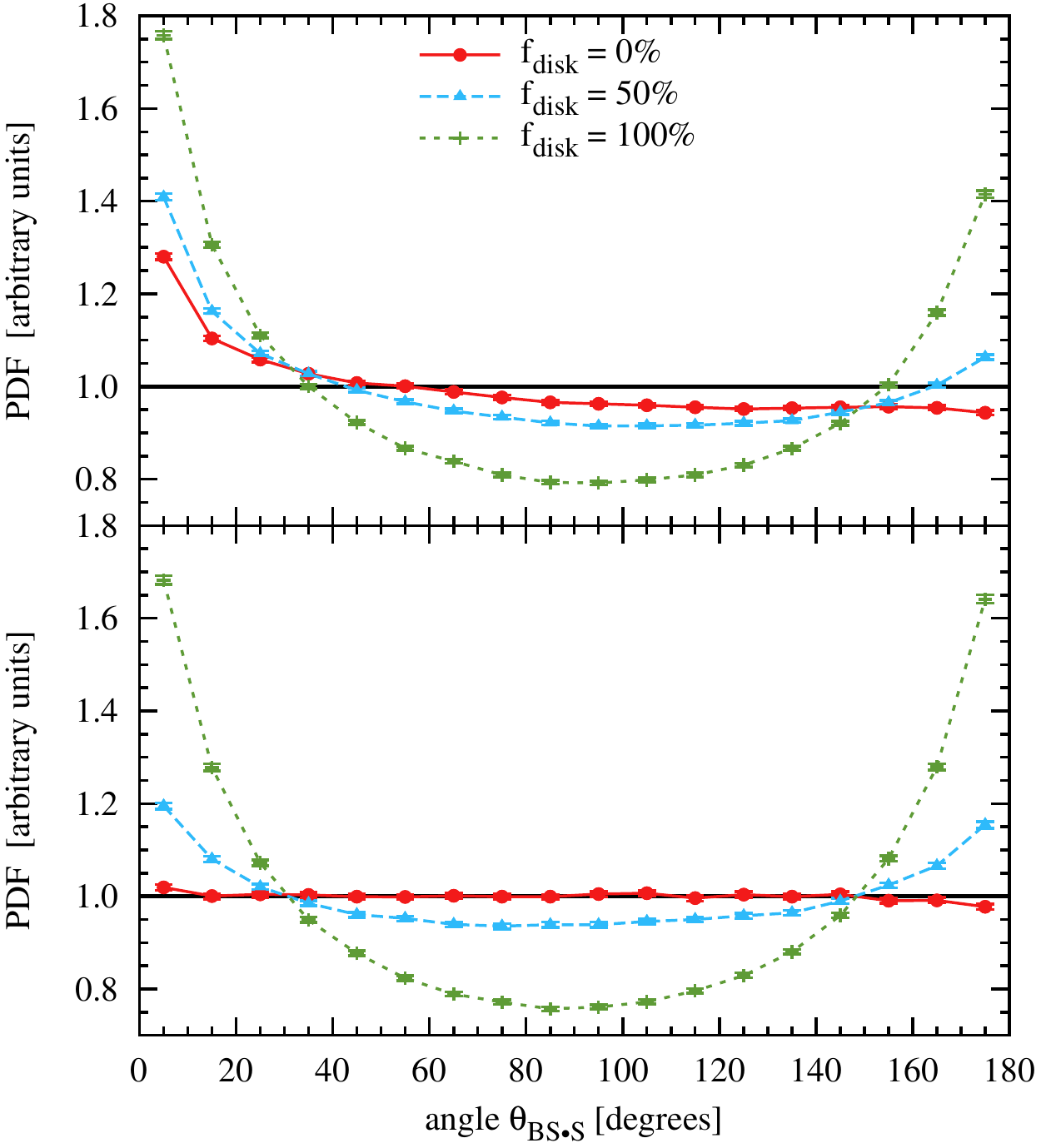}
     \caption{\textit{Top panel}: the probability distribution
       function (PDF) of the angle, $\thetaBS$, of satellites with
       respect to the line joining the primary to the brightest
       satellite in our simplified disc models. In these models a
       fraction of satellites, $f_{disc}$, of $0\%$, $50\%$ and
       $100\%$ are randomly assigned to discs, and the remaining are
       isotropically distributed around the primary
       galaxy. \textit{Bottom panel}: as above, but only for the case
       where the brightest satellite is part of the same FOF group as
       the central galaxy. }
     \label{fig:disc_model_2}
\end{figure}

To illustrate our approach we use the \MI{} data to construct a simple
model where a fraction, $f_\rmn{disc}$, of the satellites is 
distributed on a disc of $30\kpc$ thickness while the remaining
satellites are distributed isotropically. We first identify the
`friends-of-friends' (FOF) group to which each primary galaxy
belongs. Members of the FOF group other than the primary are
then randomly assigned to be part of the disc or of the isotropic
population, depending on the value of $f_\rmn{disc}$. Finally, the
satellites are spatially rearranged into the disc and isotropic
populations such that they have the same radial distribution with
respect to the primary as in the undisturbed case.

We create new mock catalogues using these disc models, which we
analyse in the same way as the SDSS data. The resulting angular
distribution function of satellites is shown in the top panel of
\reffig{fig:disc_model_2}, where the y-scale is chosen such that a
uniform distribution would correspond to a value of~1. Comparing
models with different disc fractions, it is clear that the
distribution of angles is sensitive to anisotropy in the satellite
spatial distribution. The first striking result is the asymmetry
between the $\thetaBS=0^\circ$ and $\thetaBS=180^\circ$ points, which
is unexpected given that, by construction, the satellite distributions
have cylindrical symmetry. The asymmetry is due to clustering around
interloper galaxies, which although not part of the same FOF group as
the primary, are close enough in redshift ($c|\Delta z| < 300 \kms$)
to be identified as satellites according to our selection criteria. To
quantify this effect we repeat the analysis requiring that the
brightest satellite of each primary (which defines the reference axis)
be part of the same FOF group. The result is displayed in the bottom
panel of \reffig{fig:disc_model_2} and shows that the curves in
this case are symmetric around $\thetaBS=90^\circ$.

In addition to clustering around interloper galaxies, there is also
clustering around the brightest satellite within an FOF group. This
latter effect is not captured in our simplified model where the
azimuthal angles are randomized, hence the symmetric curve in the
bottom panel of \reffig{fig:disc_model_2}. In the real case we expect
this additional clustering around the brightest satellites to 
enhance further the asymmetry of the angular distribution of satellites 
above that seen in the top panel of \reffig{fig:disc_model_2}.

The effect of clustering around the brightest satellite is
particularly evident for $\thetaBS<90^\circ$, which suggests that we
should use the $\thetaBS>90^\circ$ part of the curve for quantifying
anisotropy. For example, the disc model with $f_\rmn{disc}=0\%$ shows
a nearly flat curve for $\thetaBS>90^\circ$, as expected for an
isotropic distribution. In contrast, the model with
$f_\rmn{disc}=50\%$ shows $16\%$ more satellites at
$\thetaBS=180^\circ$ than at $\thetaBS=90^\circ$. The difference
between the values for the two angles increases to $86\%$ for
$f_\rmn{disc}=100\%$. This suggests that, with good statistics, i.e. a
large enough sample of primaries with at least one spectroscopic
satellite, the method can easily quantify the average spatial
anisotropy of the satellite distribution. Compared to previous studies
(discussed in \refsec{sec:introduction}), our analysis has the
advantage that it is independent of the correlation between the light
distribution of the primary galaxy and the anisotropy of the satellite distribution.

\subsection{The angular distribution of satellites in the SDSS and mock catalogues}
\label{sec:spatial_distribution:data}

\begin{figure}
     \centering
     \includegraphics[width=\linewidth,angle=0]{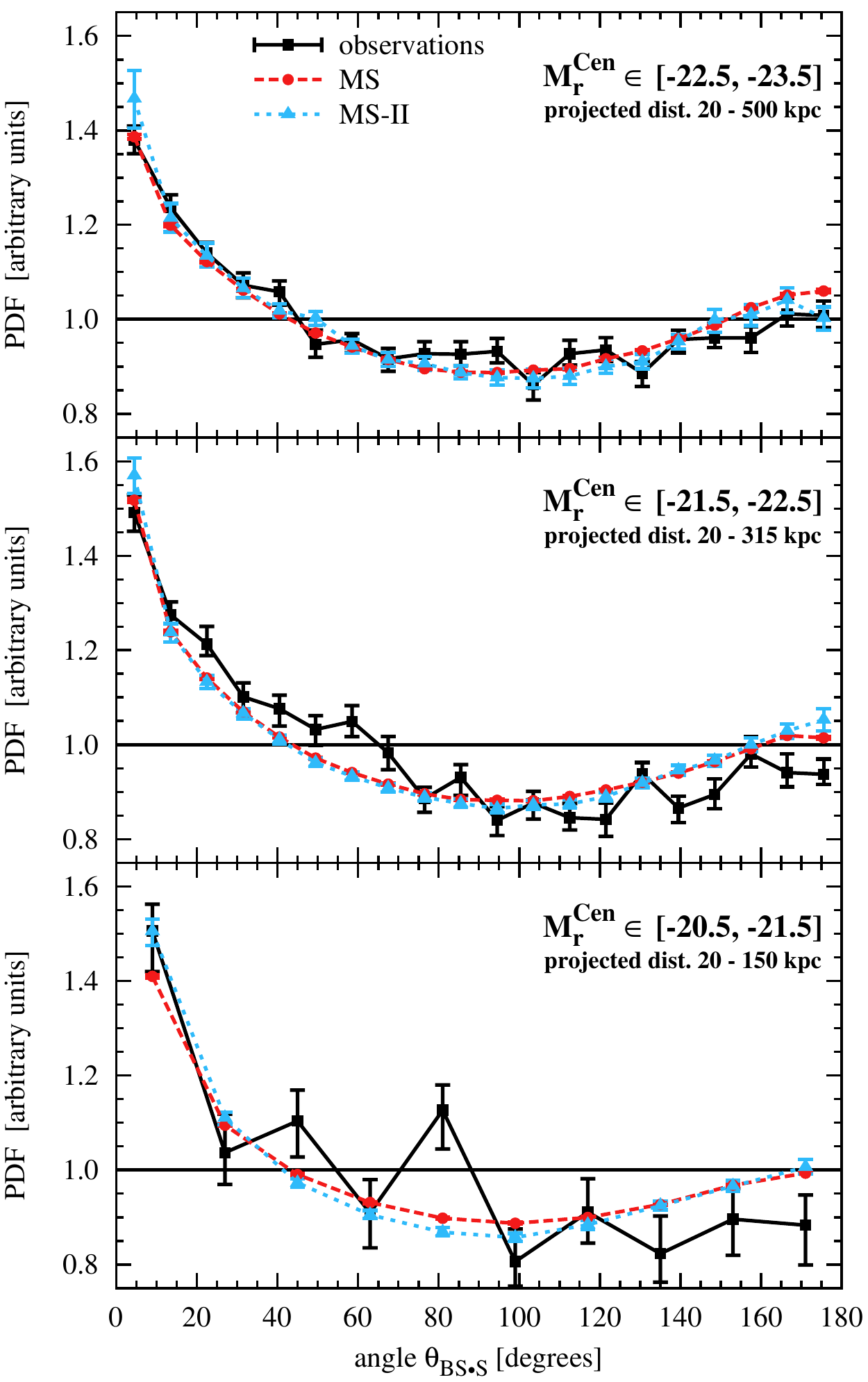}
     \caption{The probability distribution function (PDF) of the
       angle, $\thetaBS$, of satellites with respect to the line
       joining the primary to the brightest satellite. Results are
       shown for primaries in three magnitude ranges: $-22.5\ge
       M_r^\rmn{Cen}\ge-23.5$ (top), $-21.5\ge M_r^\rmn{Cen}\ge-22.5$
       (centre) and $-20.5\ge M_r^\rmn{Cen}\ge-21.5$ (bottom). The
       solid black curve is for the observational data, while the red
       and blue curves are for the \MI{} and \MII{} respectively. }
     \label{fig:projection_data}
\end{figure}

\begin{figure}
     \centering
     \includegraphics[width=\linewidth,angle=0]{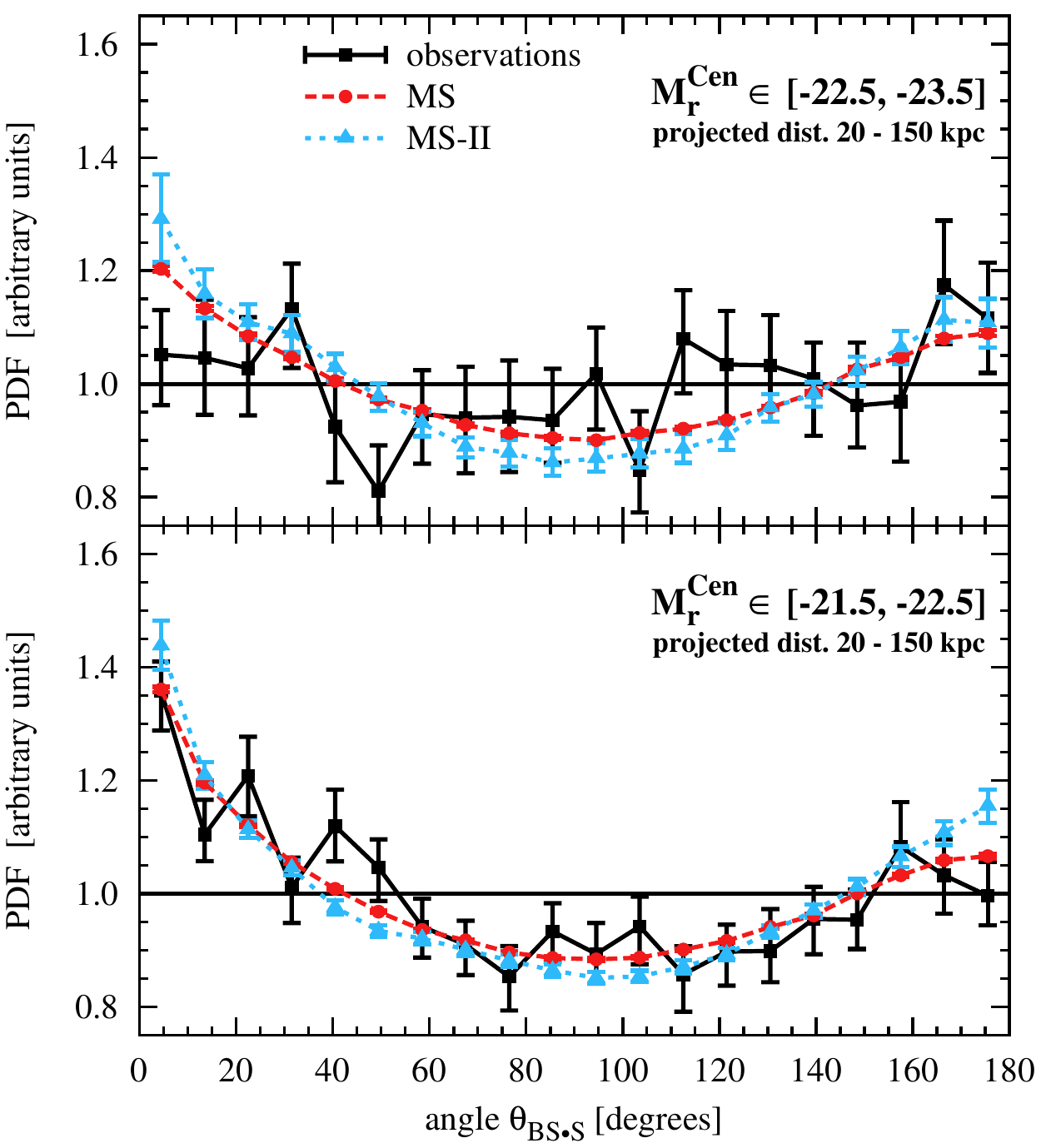}
     \caption{ \MCn{Same as \reffig{fig:projection_data} but restricted to satellites within a projected radial distance of 20 to 150$\kpc$.} }
     \label{fig:projection_data_150kpc}
\end{figure}

The angular distribution of satellites in the SDSS is given in
\reffig{fig:projection_data}, with each panel showing the results for
a different range of primary magnitudes. We use bootstrap resampling
to estimate independently errors for both observational and mock data. For the
latter, this accounts for the fact that the same object can be seen
multiple times along different sightlines. With the exception of the
faintest primary sample, where the errors are comparable to
the signal, the data clearly exhibits the telltale sign of an anisotropic
distribution: more objects at $\thetaBS=180^\circ$ than at
$\thetaBS=90^\circ$. 

Considering first the results from the mock catalogues, it is
reassuring that the \MI{} and \MII{}, which differ in mass resolution
by a factor of 125, give consistent data. This suggests that our
results are unaffected by resolution effects or by the treatment of
orphan galaxies (i.e. satellites whose dark matter halos have been
stripped). The only notable, although small, difference between the
two simulations is in the faintest magnitude bin where most of the
signal is due to satellites with $M_\rmn{r}\gsim -16$ which are not
properly resolved in the \MI{}. We have also tested the effect of
excluding orphan galaxies from the analysis and find that, in the case
of the \MII{}, the results hardly change.

In general, we find good agreement between the data and the model
  predictions. The largest deviations are seen in the central panel of
  \reffig{fig:projection_data} and are likely caused by the correlated
  deviations among the data points. The area under each PDF is the
  same, so an excess at one angle leads to a deficit at another. In
  addition, current semi-analytic models are not able to provide
  a particularly accurate match to the observed radial and colour
  distributions of satellites \citep[e.g.][]{WangW2014}. Thus, the
  small differences between data and mocks seen in
  \reffig{fig:projection_data} \MCnn{may be indicative of
  inadequacies in the semi-analytic models rather than in \lcdm{}
  itself and are not a concern for the current study.}

\MCn{In \reffig{fig:projection_data_150kpc} we investigate the spatial
  anisotropy of satellites in the projected radial range $20-150\kpc$,
  in which \ibata{} claim that ${\sim}50\%$ of satellites form
  rotating discs. We find again good agreement in the spatial
  distribution of satellites between data and mocks. In addition, the
  two mocks, from \MI{} and \MII{}, show a reasonable correspondence,
  although not as good as in \reffig{fig:projection_data}. This is
  very likely due to the treatment of orphan galaxies which, so close
  to the primary, account for most of the \MI{} satellites
  \citep[e.g. see][]{WangW2014}. }

The main conclusion from
\reffigs{fig:projection_data}{fig:projection_data_150kpc} is that the
SDSS data agree well with the results from mocks based on a
semi-analytic model of galaxy formation in \lcdm{}. \MCn{This is in
  contrast with recent claims of a conflict between the observed
  spatial anisotropy in the satellite distribution and the \lcdm{}
  model \citep[e.g.][]{Kroupa2012} . At least according to the test we
  have performed here, there is no such conflict.}  In fact, as
emphasized amongst others by \cite{Libeskind2005} and \cite{Wang2013},
spatial anisotropies are actually {\em expected} in \lcdm{}.

The simulations predict $20\%$ more satellites at $\thetaBS=180^\circ$
than at $\thetaBS=90^\circ$ for the two brightest bins, and $17\%$
more for the faintest primary sample. If we were to interpret these
results in the light of the simplified disc model introduced in
\reffig{fig:disc_model_2}, this would suggest that, on average, around
${\sim}50\%$ of the satellites are in a relatively thin
plane. \MCn{Other studies based on cosmological simulations
  \citep[eg.][]{Libeskind2005,Wang2013} showed that planes of
  satellites exist and, thus, it is natural to expect that the signal
  seen in \reffig{fig:projection_data} is related to that phenomenon.
  In cosmological simulations, these planar structures} arise from
the anisotropic infall of satellite galaxies along filaments, which
leads to the formation of flattened, pancake-like satellite
distributions \citep{Libeskind2005}.


\section{The rotation of planar structures}
\label{sec:velocity_distribution}

\begin{figure}
     \centering
     \includegraphics[width=\linewidth,angle=0]{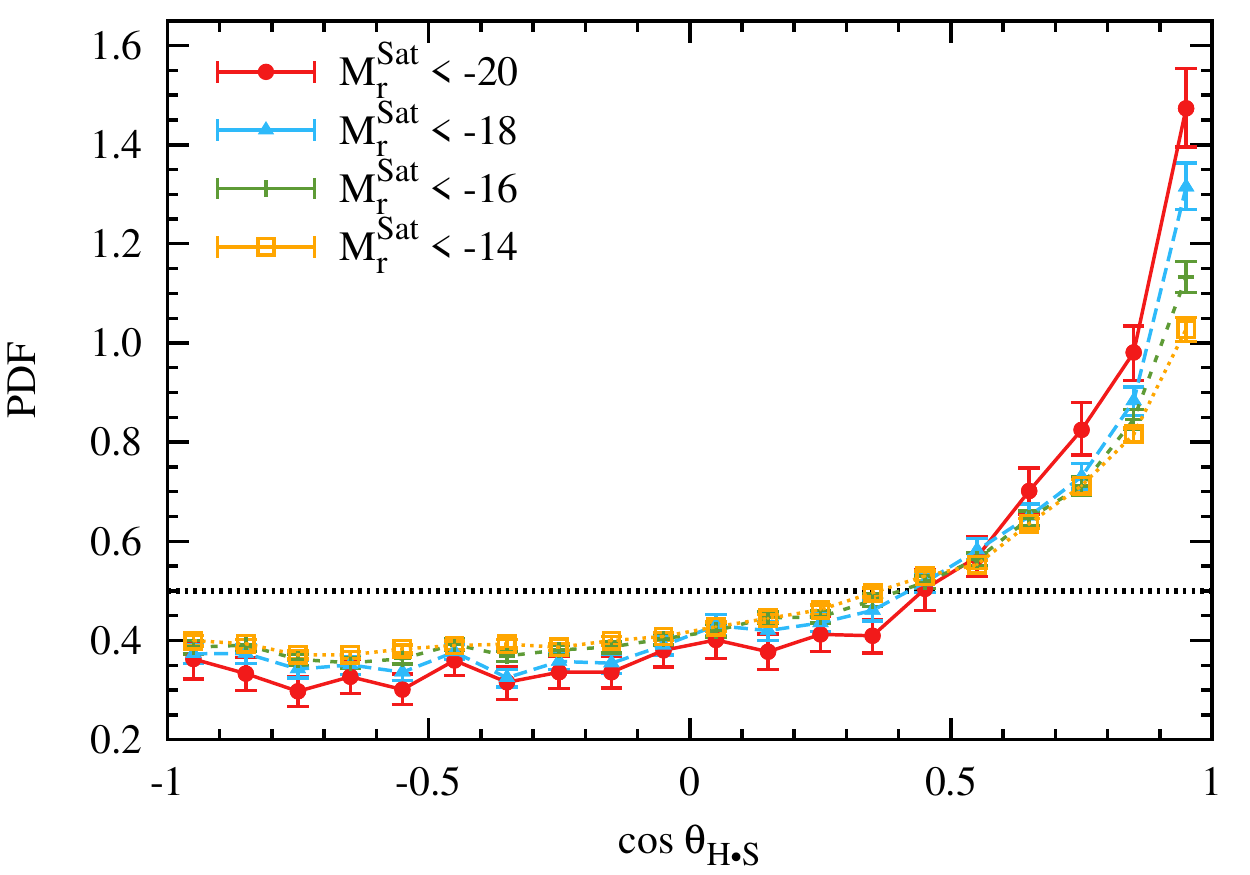}
     \caption{ The probability distribution in the \MII{} simulation
       of $\cos\thetaHS$, the cosine of the angle between the host halo spin
       and the orbital momentum of its satellites. The black
       horizontal line corresponds to a uniform distribution.  }
     \label{fig:orbit_vs_halo_spin}
\end{figure}

The motion of satellites around their primary galaxy, as predicted by
\lcdm{}, is not random, but retains a signature of their anisotropic
infall \citep{Lovell2011}. This is illustrated in
\reffig{fig:orbit_vs_halo_spin}, where we show the PDF
 of $\cos\thetaHS$, the cosine of the angle
between the halo spin and the orbital momentum of the
satellites. These results were obtained by analysing the \MII{} real
space data for central galaxies in the magnitude range $-20\ge
M_\rmn{r}^\rmn{Cen} \ge -23$. They demonstrate that satellite galaxies
rotate preferentially in the same direction as their host halo,
corroborating the results of \citet{Lovell2011}, who analysed the six
Milky Way mass halos of the Aquarius project \citep{Springel2008}. The
correlation is strongest for the brightest satellite galaxies. For
this sample, the same sense of rotation is shared, on average, by 3
times more satellites than expected from a random distribution. The
figures indicates that ${\sim}15\%$ of the satellites share the same
direction of rotation to within $25^\circ$, i.e. $\cos\thetaHS \ge
0.9$. While this represents a significant fraction of the population,
it falls shorts of the ${\sim}50\%$ fraction found in the SDSS by
\ibata{}.

\begin{figure}
     \centering
     \includegraphics[width=\linewidth,angle=0]{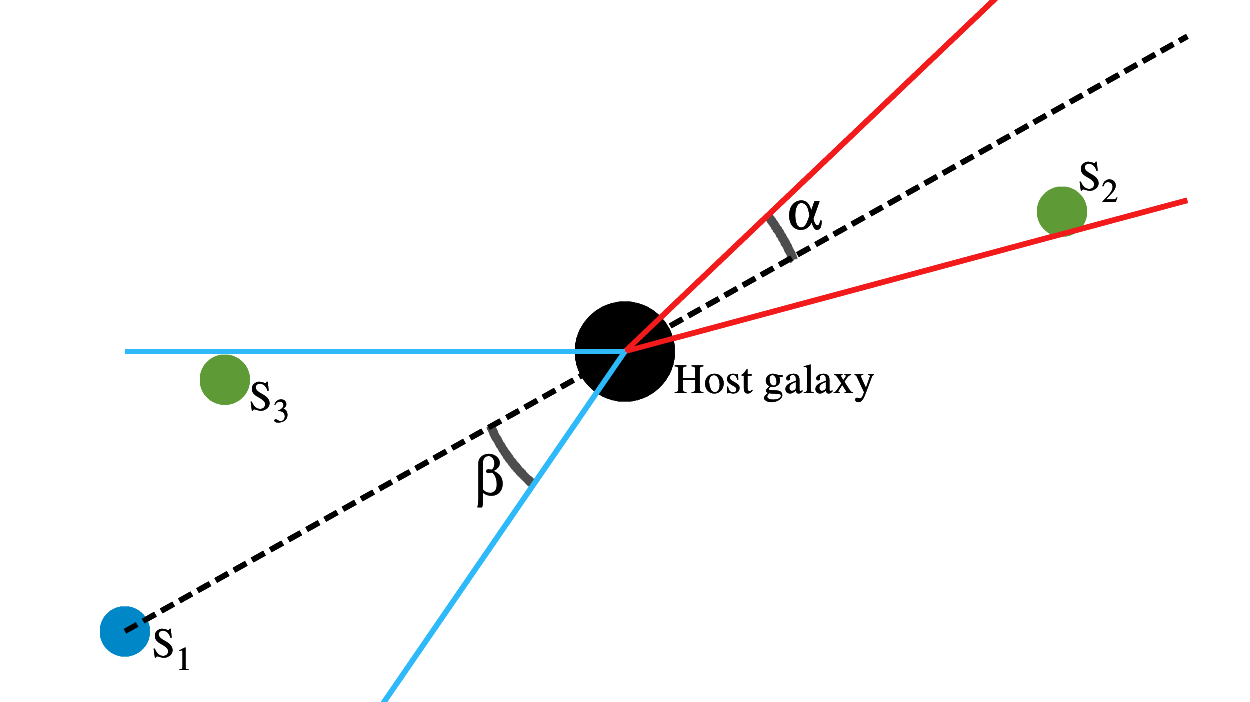}
     \caption{ Definition of the tolerance angles, $\alpha$ and
       $\beta$, used to characterise satellite pairs on opposite sides
       ($S_1S_2$) and on the same side ($S_1S_3$) of the primary galaxy,
       respectively. }
     \label{fig:tolerance_angle}
\end{figure}

\begin{figure}
     \centering
     \includegraphics[width=\linewidth,angle=0]{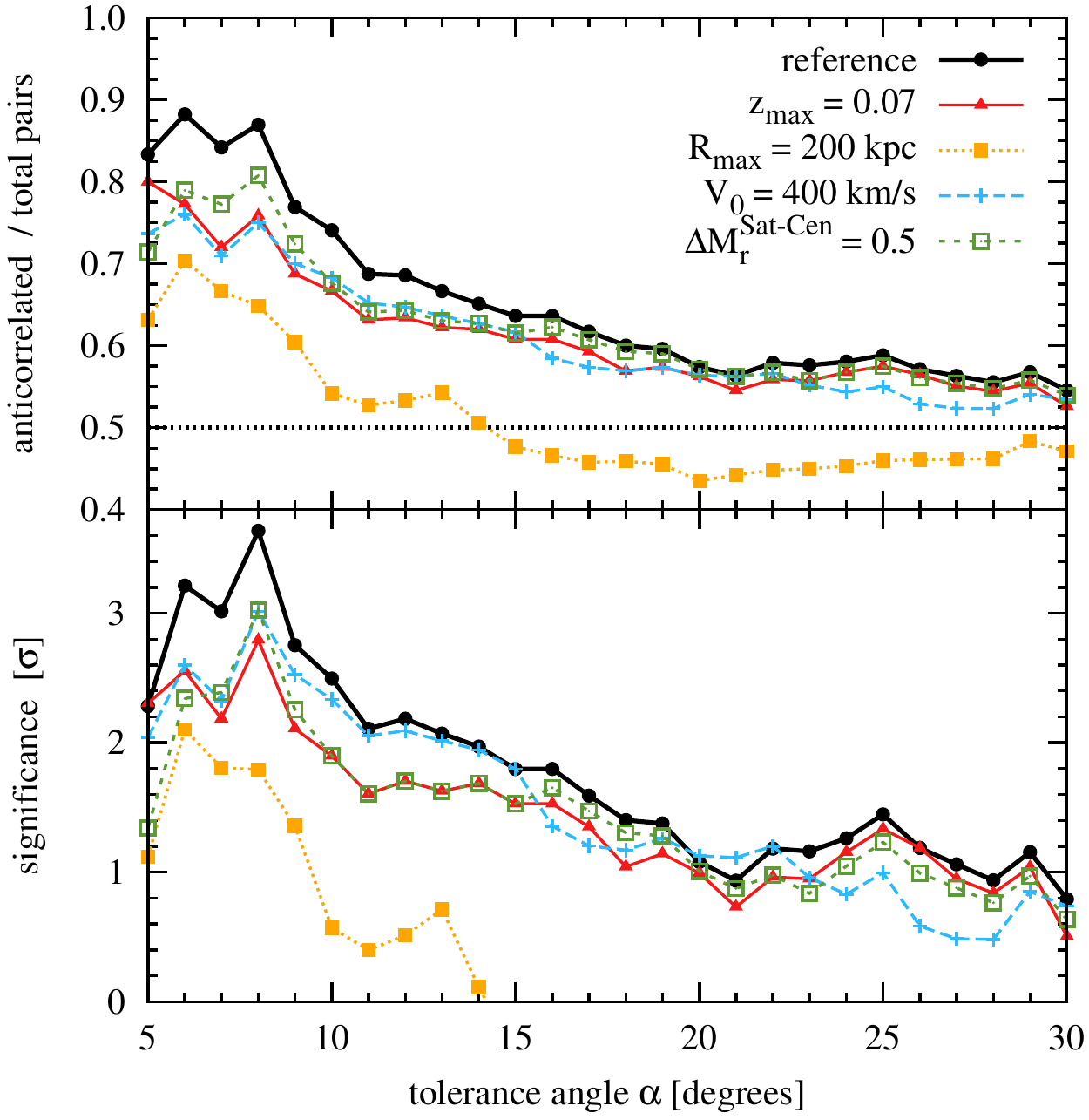}
     \caption{ \textit{Top panel}: the excess of satellite pairs with
       anticorrelated velocities as a function of the tolerance angle,
       $\alpha$, for diametrically opposite pairs. The colour curves
       show differences from the reference result (solid black) when
       varying the sample selection criteria, one at a time. Following
      \ibataTwo{}, the reference result assumes $z_\rmn{max}=0.05$,
        $R_\rmn{max}=150\kpc$, $V_{0}=300\kms$ and $\Delta
       M_r^\rmn{Sat-Cen}=1$. \textit{Bottom panel}: the significance
         of the excess of satellite pairs with anticorrelated
         velocities compared to the null hypothesis of equal numbers
         of pairs with correlated and anticorrelated velocities.  }
     \label{fig:disc_velocity_signature}
\end{figure}

\begin{figure}
     \centering
     \includegraphics[width=1.04\linewidth,angle=0]{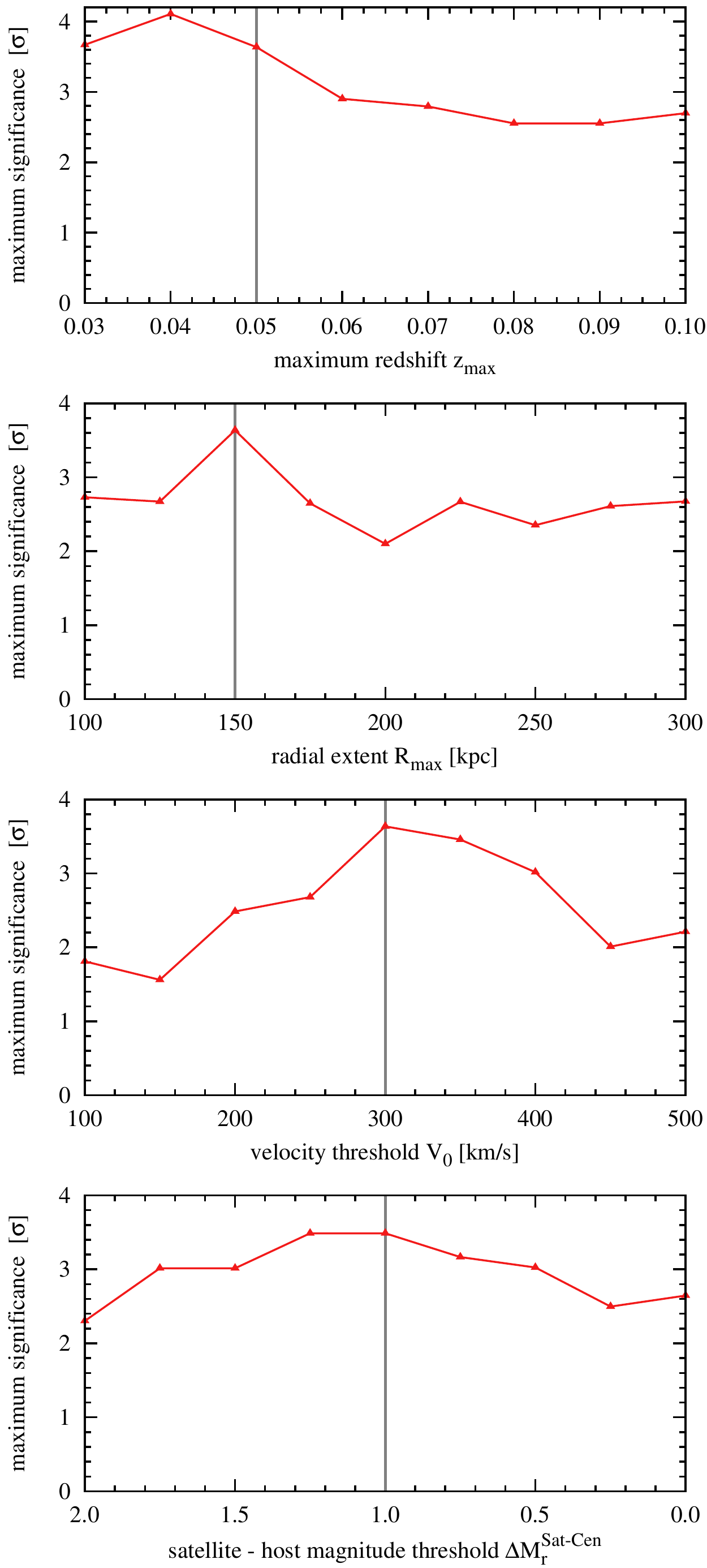}
     \caption{The maximum significance of the excess of satellite
       pairs with anticorrelated velocities 
       over the tolerance angle range $5^\circ\le \alpha \le 30^\circ$, as a function of
       different sample selection criteria. For all cases we retain the
       same selection parameters as in the reference case of \ibataTwo{}
       but vary, in turn, the maximum redshift (top panel), the radial
       extent of the volume over which satellites are found (second panel), the maximum velocity
       difference with respect to the primary  (third panel), and the
       magnitude difference between primary and satellite (bottom
       panel). The grey lines indicate the choices made in the
       reference model of \ibataTwo{}. \MCn{In each panel, the sample size increases from left to right.}
     }
     \label{fig:disc_velocity_vary_selection}
\end{figure}

To investigate the reported discrepancy between observations and
\lcdm{} predictions, we have reanalysed the SDSS data used by \ibata{}
and extended this kind of analysis in order to obtain better
statistics.  We are interested in the number of satellite pairs with
correlated and anticorrelated line-of-sight velocities as a function
of a tolerance angle that characterises the angular separation of the
pair, as illustrated in \reffig{fig:tolerance_angle}. The tolerance
angle, $\alpha$, refers to satellite pairs on diametrically opposite
sides of the primary, following the same convention as \ibata{}, while
the tolerance angle, $\beta$, refers to pairs on the same side of the
host.

\subsection{Diametrically opposed satellite pairs}
\label{subsec:velocity_distibution_opposed}
To begin with our analysis follows the exact sample selection criteria
described by \ibata{} (see \refsec{sec:data} for details). At a
tolerance angle, $\alpha=8^\circ$, we were able to recover only 20
pairs of diametrically opposite satellites compared to the 22 pairs
reported by \ibata{}. This discrepancy is likely due to the use of
different versions of the NYU-VAGC catalogue\footnote{There is an
  ambiguity regarding the catalogue used by \ibata{} since the
  NYU-VAGC website they referenced contains a multitude of
  catalogues. After trying several of them, we settled on the one
  which gives absolute magnitudes closest to the values given in
  Table~1 of \ibata{}. Nevertheless, there is a ${\sim}0.03$ scatter
  between the absolute magnitudes in our catalogue and those quoted by
  \ibata{}.}. Our original sample missed two pairs with anticorrelated
velocities that appear in the sample of \ibata{}, corresponding to
rows 1 and 18 in their Table~1. Using VizieR, we found the satellite
pair in row 18 in another catalogue, but we could not identify one of
the satellites of the pair in row 1. Nevertheless, we have chosen to
include both these pairs in our sample. We also found an additional
pair with $\alpha<8^\circ$, which has correlated velocities, that does
not appear in the \ibata{} sample.

\begin{table*}
    \centering
    \caption{ The fraction of diametrically opposed pairs with anticorrelated velocity and the significance of the excess when varying the sample selection criteria. We give the results for a tolerance angle, $\alpha=8^\circ$, that corresponds to the maximum significance shown in \reffig{fig:disc_velocity_vary_selection}. The column in grey corresponds to the reference selection criteria of \ibataTwo{}. } 
    \label{tab:disc_velocity_pair_count}
    \begin{tabular}{L{3cm}L{4.cm} rrG rrr rr }
        \hline
        \multirow{3}{*}{$z_\rmn{max}$ variation}   
                                         & $z_\rmn{max}$ values         &  0.03  &  0.04 &  0.05 &  0.06 &  0.07 &  0.08 &  0.10  \\
                                         & anticorrelated / total pairs & 10/10  & 18/19 & 20/23 & 21/27 & 22/29 & 22/30 & 23/31  \\
                                         & significance [$\sigma$]      & 3.7    & 4.1   & 3.6   & 2.9   & 2.8   & 2.6   & 2.7    \\
        \hline
        \multirow{3}{*}{$R_\rmn{max}$ variation}   
                                         &  $R_\rmn{max}$ values        &  100   &  125  &  150  &  175  &  200$^\dagger$  &  250$^\dagger$  &  300$^\dagger$  \\
                                         & anticorrelated / total pairs & 10/11  & 14/17 & 20/23 & 21/28 & 24/37 & 38/65 & 48/80  \\
                                         & significance [$\sigma$]      & 2.7    & 2.7   & 3.6   & 2.6   & 1.8   & 1.3   & 1.8    \\
        \hline
        \multirow{3}{*}{$V_\rmn{max}$ variation}   
                                         &  $V_\rmn{max}$ values        &  200   &  250  &  300  &  350  &  400  &  450  &  500   \\
                                         & anticorrelated / total pairs & 11/13  & 12/14 & 20/23 & 23/28 & 27/36 & 31/48 & 34/52  \\
                                         & significance [$\sigma$]      & 2.5    & 2.7   & 3.6   & 3.5   & 3.0   & 2.0   & 2.2    \\
        \hline
        \multirow{3}{*}{$\Delta M_r^\rmn{Sat-Cen}$ variation}   
                                         &  $\Delta M_r^\rmn{Sat-Cen}$ values  &  1.50  &  1.25 &  1.00 &  0.75 &  0.50 &  0.25 &  0.00   \\
                                         & anticorrelated / total pairs        & 17/20  & 20/23 & 20/23 & 20/24 & 21/26 & 21/28 & 22/29  \\
                                         & significance [$\sigma$]             & 3.2    & 3.6   & 3.6   & 3.3   & 3.2   & 2.6   & 2.8    \\
        \hline
        \multicolumn{10}{p{.88\textwidth}}{$^\dagger$ The $R_\rmn{max}=200$, 250 and $300\kpc$ entries have the maximum excess of anticorrelated pairs for $\alpha=6^\circ$, for which there are 19/27, 26/36 and 31/46 anticorrelated pairs that correspond to a 2.1, 2.4 and 2.7$\sigma$ excess. At $\alpha=6^\circ$, the reference sample has 15/17 anticorrelated pairs.} \\
    \end{tabular}
\end{table*}

The excess of pairs with anticorrelated velocities and its
significance as a function of the tolerance angle, $\alpha$, is shown
by the thick black line in \reffig{fig:disc_velocity_signature}. The
significance of the excess is evaluated as the sigma-value
corresponding to the probability of obtaining such an excess for a
binomial distribution of mean $0.5$. The inclusion of an additional
pair with correlated velocities in our sample results in a smaller
excess of anticorrelated pairs than found by \ibata{} and a
correspondingly lower statistical significance. The most significant
excess is found at $\alpha=8^\circ$ and corresponds to a $3.6\sigma$
significance, compared to a maximum significance of $4\sigma$ reported
by \ibata{} at the same tolerance angle. 

\reffig{fig:disc_velocity_signature} also shows how the excess of
anticorrelated velocity pairs changes when the sample selection
criteria are relaxed. We vary one parameter of the selection criteria
at a time, keeping the remaining parameters at their reference values
as given in \refsec{sec:data}. In all cases we find that the excess of
anticorrelated pairs decreases as does the corresponding maximum 
significance of the excess. 

We explore further the sensitivity of the excess of anticorrelated
velocity pairs by systematically varying, one at a time, some of the
parameters used to select the sample. In each case, we determine the
maximum significance of the signal over the range of tolerance angles,
$5^\circ\le \alpha \le 30^\circ$. With few exceptions, the maximum
significance is found for $\alpha=8^\circ$. The maximum significance
as function of some of the main parameters in the selection criteria
is plotted in \reffig{fig:disc_velocity_vary_selection}. For clarity,
the reference values for each parameter are shown as a vertical grey
line. We find that small variations in the sample selection parameters
can lead to a significant reduction in the significance of the
observed excess of anticorrelated velocity pairs.  Except for a few
values, the maximum significance is below the $3\sigma$
level. 

In \reftab{tab:disc_velocity_pair_count} we list the total number
of pairs and the number of pairs with anticorrelated velocities for
the samples plotted in \reffig{fig:disc_velocity_vary_selection}.
\MCnn{If we tighten the selection criteria, the signature of rotating satellite systems is still present, though in most cases its significance is reduced, since the resulting sample is a subset of the reference sample.}
Relaxing the $z_\rmn{max}$ or the $\Delta M_r^\rmn{Sat-Cen}$ selection criteria
  adds at most a small number of new pairs. Increasing $z_\rmn{max}$
  from $0.05$ to $0.1$ adds $8$ extra pairs, while decreasing $\Delta
  M_r^\rmn{Sat-Cen}$ from $1\;\rmn{mag}$ to $0\;\rmn{mag}$ adds $6$
  additional pairs. So, for these cases, the measurement is always
  dominated by the 23 pairs found in the reference sample. In
  contrast, relaxing $R_\rmn{max}$ and $V_\rmn{max}$ adds
  significantly more pairs. Increasing $R_\rmn{max}$ from $150$ to
  $300\kpc$ adds 57 new pairs. Out of these, only 28, exactly half the
  sample, have anticorrelated velocities. Similarly, increasing
  $V_\rmn{max}$ from $300$ to $500\kms$ adds 29 new pairs, with 14 of
  them, again half the sample, having anticorrelated velocities. Thus,
  there is no signature of a rotating disc for $V_\rmn{max}\ge300\kms$
  or for $R_\rmn{max}\ge150\kpc$. Any large excess of pairs with
  anticorrelated velocities seen in
  \reffig{fig:disc_velocity_vary_selection} is therefore entirely
  driven by the reference sample, since the measurements are not independent: 
  they all contain most or all of the 23
  pairs of the default sample.

The choices made by \ibata{} reflect various compromises (R. Ibata,
private communication). The maximum redshift cut, $z_\rmn{max}=0.05$,
was chosen because this value has been commonly used in similar
studies to avoid including very bright satellites. The search radius,
$R_\rmn{max}=150\kpc$, was chosen to match the M31 PAnDAS survey,
while the velocity threshold, $V_\rmn{max}=300\kms$, corresponds to
twice the central velocity dispersion of Andromeda. The maximum
magnitude difference, $\Delta M_r^\rmn{Sat-Cen}=1$, between satellites
and the central galaxy was chosen in order to discard objects that are
too close in brightness to the host. 

These choices, of course, are to some extent arbitrary. For example, 
increasing the maximum redshift range from $z_\rmn{max}=0.05$ to
$0.07$ adds mainly bright primaries with absolute r-band magnitudes in
the range $[-22.6,-22.0]$ which already includes more than half of the
primaries in the reference sample.  Similarly, increasing the maximum
radius used to identify satellites from $R_\rmn{max}=150$ to
$200\kpc$ is not unreasonable given that most of the galaxies in our
sample occupy halos with a virial radius larger than $200\kpc$.  The
sensitivity of the results to the details of the sample selection lead
us to conclude that the detection of systemic rotation in the
satellite population with current observational samples is not robust.

\subsection{Same-side satellite pairs}

\begin{figure}
     \centering
     \includegraphics[width=1.0\linewidth,angle=0]{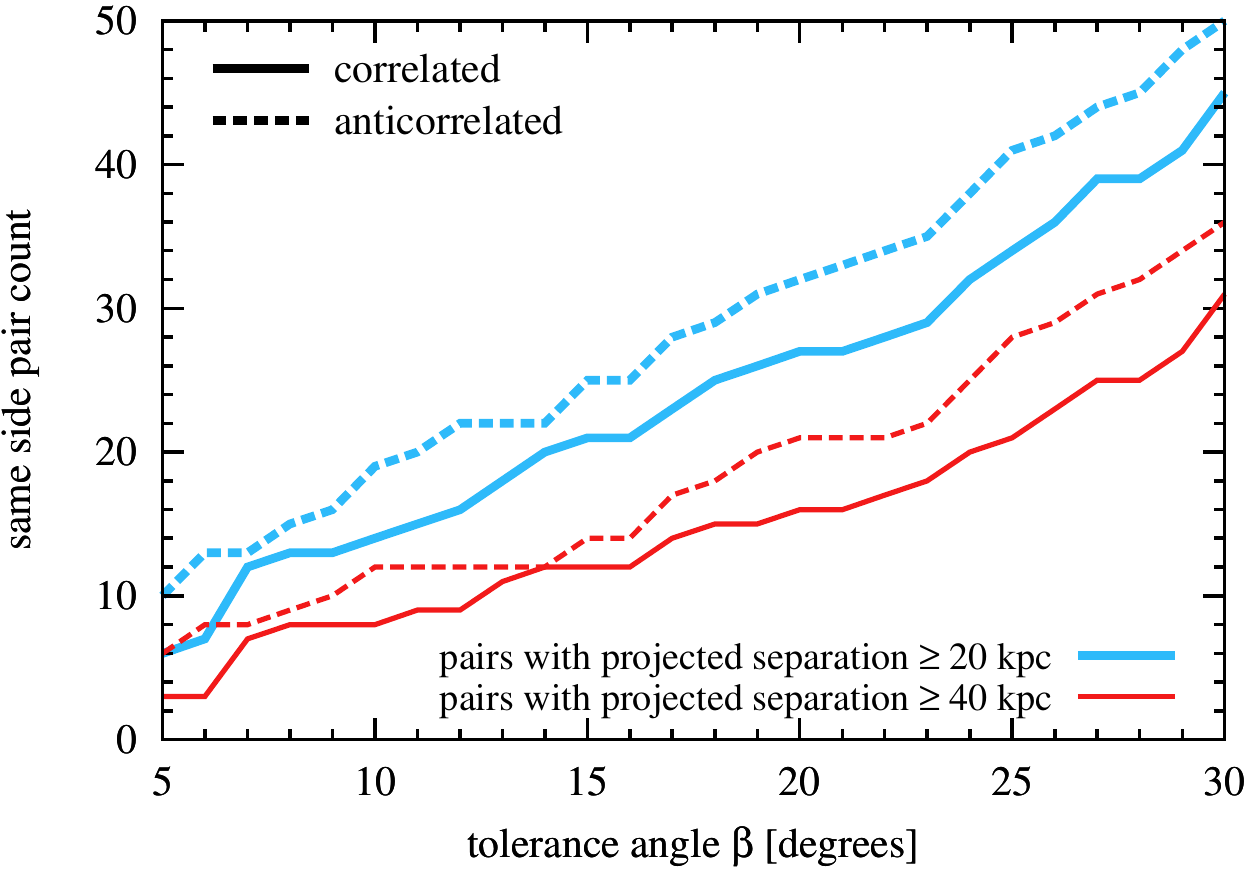}
     \caption{ \MCn{The number of same side satellite pairs with correlated and anticorrelated velocities as a function of the tolerance angle, $\beta$. We present results for satellite pairs separated in projected distance by more than $20\kpc$ (thick lines) and more than $40\kpc$ (thin lines).} }
     \label{fig:disc_velocity_same_side_pair_count}
\end{figure}

If the excess of pairs with {\it anticorrelated} velocities on
opposite sites of the host were attributable to rotating discs of
satellites, an equal but opposite excess of {\it correlated}
velocities would be expected for pairs of satellites on the same side
of the primary. This provides an independent test of the
significance of the result reported in the previous section. For
same-side satellite pairs we use the same selection criteria as
described in \refsec{sec:data} and in \ibata{}, but also require the
projected distance between satellites to be greater than $20\kpc$. 
\MCn{In the SDSS sample, due to fibre collisions, very few satellite
  pairs are closer than this separation, so we require projected
  separations $\ge20\kpc$ in order to make sure that the mocks do
  not include close pairs likely to be absent in the real data.}

\MCn{The number of same side pairs with correlated and anticorrelated
  velocities is shown in
  \reffig{fig:disc_velocity_same_side_pair_count}. We find no
  significant difference in the number of pairs with correlated and
  anticorrelated velocities, within the expected scatter of a binomial
  distribution. The same result holds if we increase the minimum
  separation between pairs to $40\kpc$, although in this case the
  sample is, of course, smaller. For the remainder of this section, we
  consider pairs with a projected separation $\ge20\kpc$, to make use
  of the better statistics available for his sample. The number of
  same-side pairs at the tolerance angle, $\beta=8^\circ$, is 28,
  which is only slightly larger than the total count of 23
  diametrically opposed pairs. Therefore, same-side pairs have similar
  or better statistical power to test for the presence of
  rotating discs of satellite galaxies.}

\begin{figure}
     \centering
     \includegraphics[width=1.0\linewidth,angle=0]{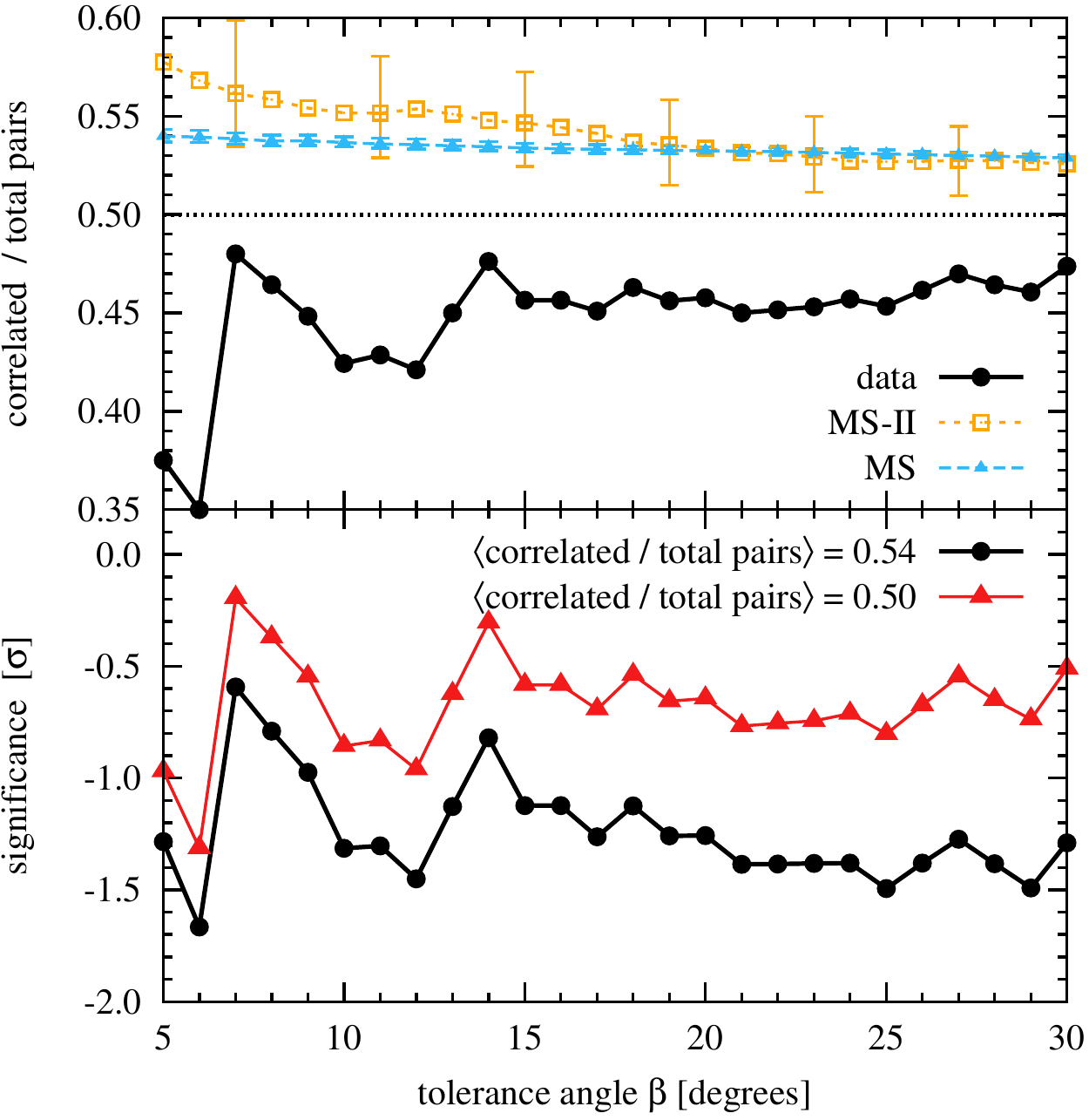}
     \caption{ \textit{Top panel}: the excess of satellite pairs with
       correlated velocities as a function of the tolerance angle,
       $\beta$, for pairs found on the same side of the primary. We
       compare the observational data (solid black) with results from
       the \MI{} (blue crosses) and \MII{} (orange open squares). The
       signature of rotating discs of satellites is an excess of pairs
       with correlated velocities. In contrast, the data show a small
       deficit of pairs with correlated velocities. The error bars for
       the \MI{} and \MII{} data show $1\sigma$ bootstrap
       uncertainties. \textit{Bottom panel}: the significance of the
       excess of satellite pairs with correlated velocities compared
       to the \MI{} predictions (circles; mean expectation of $0.54$)
       and to a uniform distribution (triangles; mean expectation of
       $0.5$). The negative values for the significance reflect the
       fact that instead of an excess, we find a deficit of correlated
       pairs.  }
     \label{fig:disc_velocity_same_side}
\end{figure}

\reffig{fig:disc_velocity_same_side} shows the fractional abundance of
correlated velocities and its significance as a function of the
tolerance angle, $\beta$, for same-side satellite pairs. Instead of
the expected excess, we find a small deficit of pairs with correlated
velocities, although the result is consistent, within $1\sigma$, with
a uniform distribution and only marginally inconsistent,
${\sim}1.5\sigma$, with the results from the mock catalogue. Even in
the absence of discs, the \MI{} and \MII{} simulations predict a
slight excess, $54\%$, of correlated velocity pairs, which may be due
to binary satellites orbiting around the brighter primary\footnote{For
  diametrically opposed pairs the simulations predict a
  approximatively equal numbers of anticorrelated and correlated
  velocity pairs.}. The idea of rotating discs of satellites is
disfavoured by the lack of excess of same-side pairs with correlated
velocities and supports the low significance of the excess of
anticorrelated velocities for opposite-side pairs found in the
preceding section.

\section{{\em Concerns raised by Ibata et al.}}
\label{sec:ibata}

\MCn{A few weeks after submitting our paper to the {\em arXiv},
  \citet[][hereafter \ibataThree{}]{Ibata2014d} posted a paper in
  which they included a response to the concerns we had expressed in
  our original submission about \ibata{}'s earlier results. In the
  remainder of this section we address their response and show that our
  initial critique of the robustness of the detection of a rotating
  disc of satellites remains valid.}

{\em 1. Spatial distribution}

\MCn{\ibataThree{} investigated the spatial anisotropy of satellites
  by counting, as a function of opening angle, all satellite pairs
  in which one member is a spectroscopically-confirmed satellite of
  the primary while the second is a photometric-redshift satellite
  candidate. This is very similar to the analysis we presented in
  \refsec{sec:spatial_distribution}, with the difference that
  \ibataThree{} considers all the spectroscopically-confirmed
  satellites of a primary, not only the brightest one as in our
  case. In practice, this difference is unimportant since most
  primaries have only one such satellite. Thus, their Fig. 3 is
  equivalent to the $90^\circ \le \thetaBS \le 180^\circ$ region of
  our \reffigs{fig:disc_model_2}{fig:projection_data_150kpc}, where
  their angle definition is the complement of our angle, $\thetaBS$.} 

\MCn{Due to limited statistics, \ibataThree{} focused their analysis
  on the ratio, $\NOA{}$, between the number of satellites with
  $135^\circ\le\thetaBS$ and the number with
  $90^\circ\le\thetaBS\le135^\circ$, finding a ratio of
  ${\sim}3$. This is in stark contrast to our analysis which finds a
  significantly lower value of $\NOA{} = 1.1$. In fact, a value of $3$
  is unphysical as may be easily verified using simplified models such
  as the one we introduced in
  \refsec{sec:spatial_distribution:disc_model}. We 
  generated mock catalogues from such simplified models, to which we
  applied the same primary and satellite selection criteria as 
  \ibataThree{}. The case when $50\%$ of satellites are on a thin
  plane, with the rest distributed isotropically, gives
  $\NOA{}=1.1$. Even in the most extreme case, when all the satellites
  are on an infinitely thin plane, the $\NOA{}$ ratio cannot be higher
  than ${\sim}1.4$. This is because the $\NOA{}$ ratio measures the
  anisotropies of the satellite distribution as projected on the plane
  of the sky. The anisotropy is largest for satellite planes seen
  edge-on, and decreases rapidly to zero as the viewing angle
  approaches a face-on planar configuration. The signal is further
  diminished by contamination from interloper galaxies that have a
  small line-of-sight velocity difference with respect to the primary
  and thus are mistaken as spectroscopically confirmed satellites (see
  \reffig{fig:disc_model_2} for a qualitative estimate of this
  effect).}

\MCn{We suspect that the high $\NOA{}$ ratio found by \ibataThree{} is
  due to an overestimation of the background contamination. Even a
  small change in the background count will result in a large change
  in the $\NOA{}$ value. For example, a decrease in the background
  fraction from their quoted value of $85\%$ to a more modest $80\%$
  would lower the \ibataThree{} result to $\NOA{}=2.1$. A similar, or
  even larger, decrease in background fraction is not unlikely given
  that \ibataThree{} used a new background estimation method that has
  not been tested in any systematic way. Their background contamination was
  estimated using bright satellites (apparent magnitude $r\lsim17.7$)
  with spectroscopic redshifts and then extrapolating the result to
  much fainter objects ($r\lsim19.5$). In contrast, our background
  contamination is estimated in a very robust way using a method that
  has been thoroughly and independently tested and applied by several
  groups \citep[among others
  by][\WangWhite{}]{Nierenberg2012,Guo2012}.}

  \MCn{Secondly, \ibataThree{}'s choice of radial extent used to identify
  satellites, between $100$ and $150\kpc$, was motivated by the
  presence of a peak at these radii seen in their Fig.~5. Given that for radial distances ${\lsim}
  180\kpc$ all the data points agree well within the $1\sigma$
  uncertainty, the peak is more likely to be a statistical fluctuation
  rather than a real signal.  Such {\em a posteriori} choice of the
  radial extent that maximizes the enhancement will inevitably lead to
  an excess in the $\NOA{}$ ratio due solely to statistical
  fluctuations. }

\begin{figure}
     \centering
     \includegraphics[width=1.0\linewidth,angle=0]{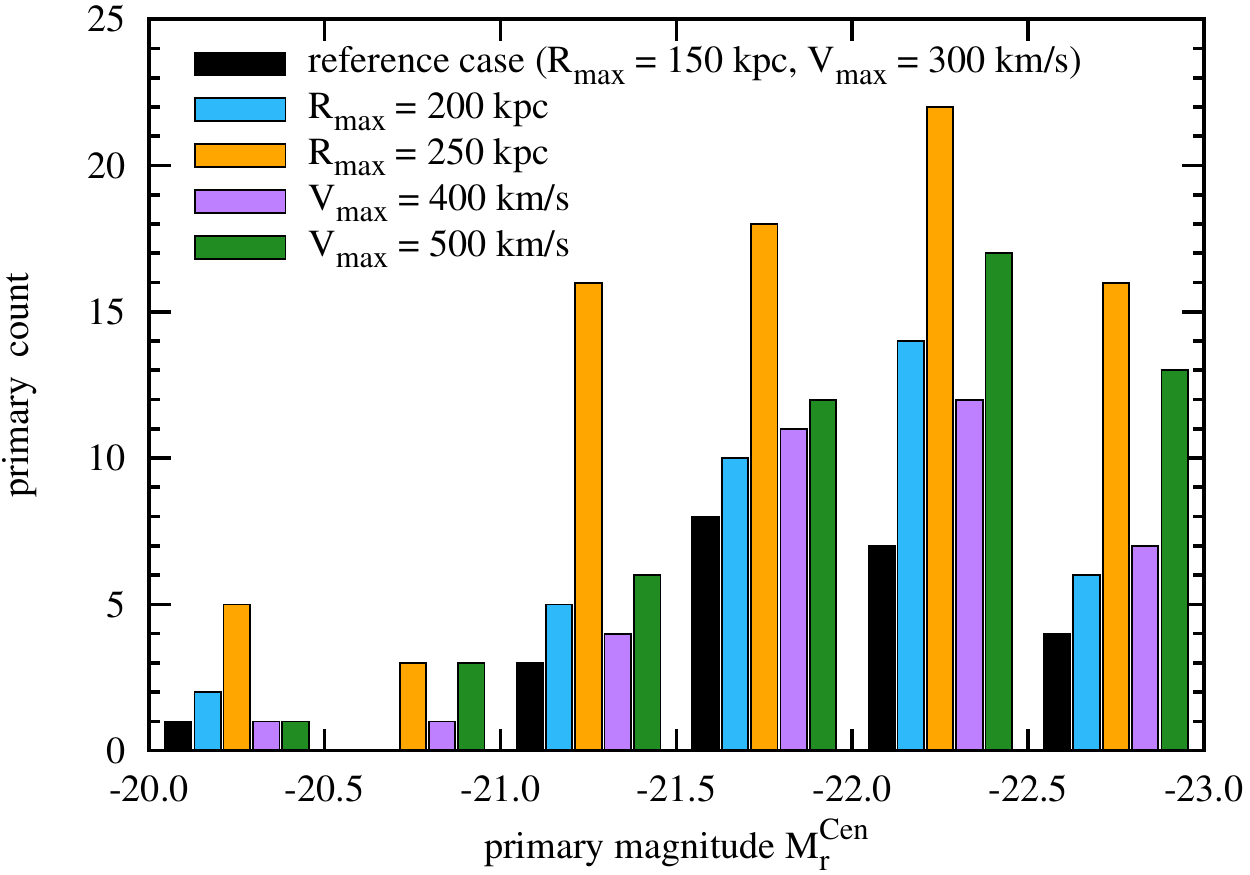}
     \caption{ \MCn{The histogram of absolute r-band magnitudes,
         $M_\rmn{r}^\rmn{Cen}$, for isolated primaries that host a
         diametrically opposed satellite pair within a tolerance
         angle, $\alpha=8^\circ$. Results are presented for several
         sample selection criteria: the reference sample as in
         \ibataTwo{}; when increasing the maximum projected radial
         extent from $R_\rmn{max}=150$ to $200$ and $250\kpc$; and
         when increasing the maximum velocity difference with respect
         to the primary from $V_\rmn{max}=300$ to $400$ and
         $500\kms$.} } 
     \label{fig:ibata_host_histogram}
\end{figure}

{\em 2. Diametrically opposed pairs}

\MCn{\ibataThree{} put forward several reasons attempting to explain
  why we find a decrease in the excess of anticorrelated pairs in
  their diametrically opposed-pair test when we vary their sample
  selection criteria. They used these arguments to downplay the
  results of \refsec{subsec:velocity_distibution_opposed} above but,
  unfortunately, they did not check if their arguments are actually
  valid in practice. For simplicity, we address their concerns
  regarding the variation of the $R_\rmn{max}$ and $V_\rmn{max}$
  selection parameters since, when relaxing these, we obtain the
  largest number of new satellite pairs (see
  \reftab{tab:disc_velocity_pair_count} and the discussion in
  \refsec{subsec:velocity_distibution_opposed}). Thus, relaxing these
  selection criteria offers the cleanest way to assess the robustness
  of the results presented in \ibata{}.}

\MCn{\ibataThree{} stated that increasing the maximum radial extent,
  $R_\rmn{max}$, or the maximum line-of-sight velocity difference,
  $V_\rmn{max}$, leads to a significant increase in contamination by
  interloper galaxies that are not true satellites of the
  primary. This appears very unlikely, since, as can be seen in
  \reffig{fig:ibata_host_histogram}, most of the primaries found in
  the original sample of \ibata{} are very bright, with virial radii
  significantly larger than $150\kpc$ and with satellite velocity
  dispersions larger than Andromeda's. Increasing $R_\rmn{max}$ or
  $V_\rmn{max}$ does not introduce any bias in the primaries around
  which satellites are found: a Kolmogorov-Smirnov test indicates that
  all five distributions of primary magnitudes shown in
  \reffig{fig:ibata_host_histogram} are consistent with one another
  at $99\%$ confidence. Thus, by increasing $R_\rmn{max}$ or
  $V_\rmn{max}$ we simply find more satellite pairs around primaries
  similar to the ones in the reference sample.}

\MCn{We can estimate the the contamination fraction due to interloper
  galaxies using our \MII{} mock catalogues which have a similar
  background to the real data. For this, we find the fraction of
  \MII{} satellite pairs in which one or both members of the pair are
  not part of the same friends-of-friends halo as the isolated
  primary. This interloper fraction is highly dependent on the
  brightness of the primary, so we compute it for each of the six
  primary magnitude bins shown in
  \reffig{fig:ibata_host_histogram}. For example, the contamination
  fraction for satellite pairs selected using the reference criteria
  around centrals with $-20.5\le M_\rmn{r}^\rmn{Cen}\le-20.0$ is
  $18\%$ and this fraction decreases to $3.5\%$ for centrals with
  $-23.0\le M_\rmn{r}^\rmn{Cen}\le-22.5$. To find the mean
  contamination for each sample, we weigh the contamination fraction
  found for each primary magnitude bin by the number of primaries in
  that bin.}

\MCn{For the reference sample we find a mean contamination fraction of
  $6\%$. When we increase $V_\rmn{max}$ from $300$ to $400$ and $500$
  $\kms$ we find the same mean contamination fraction of $6\%$.  The
  reason for this is that while the contamination fraction for each
  primary magnitude bin increases, the fraction of brighter primaries
  is slightly higher than in the reference case, so that, on average,
  the mean sample contamination hardly changes. When we increase
  $R_\rmn{max}$ from $150$ to $200$, $250$ and $300\kpc$, we find a
  slight increase in contamination from $6$ to $8$, $10$ and $12\%$
  respectively. Thus, the change in mean sample contamination is minor
  and cannot possibly explain why satellite pairs found at
  $V_\rmn{max}>300\kms$ or $R_\rmn{max}>150\kpc$ do not show any
  excess of anticorrelated over correlated velocities (see discussion
  in \refsec{subsec:velocity_distibution_opposed}).}

{\em 3. Same side pairs}

\begin{figure}
     \centering
     \includegraphics[width=1.0\linewidth,angle=0]{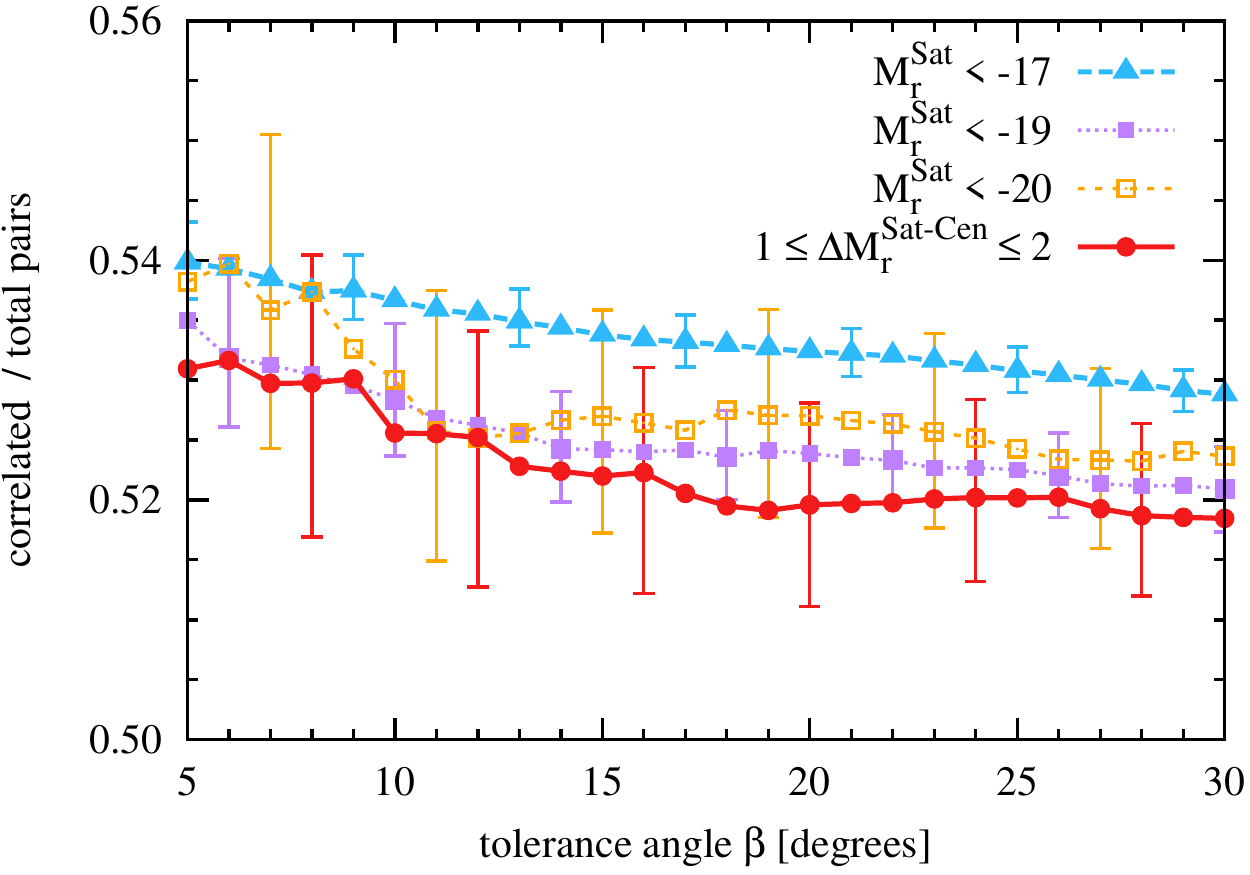}
     \caption{The excess of same-side satellite pairs with
         correlated velocities predicted in the \MI{} simulation mock
         catalogue for satellites of different luminosity. The excess
         is largely independent of satellite luminosity and even very
         bright satellite pairs, within 1 to 2 magnitudes from their
         primaries, show a similar signal. Thus, the lack of a
         signature of rotating satellite systems in the same-side
         pair test cannot be due to the presence of bright binary
         pairs. The solid triangles correspond to then blue crosses in
         the top panel of \reffig{fig:disc_velocity_same_side}. Note
         the very different scales used in the two figures.}
     \label{fig:ibata_same_side_pairs}
\end{figure}

\MCn{In \reffig{fig:ibata_same_side_pairs} we investigate if the lack
  of a rotating disc signal for same-side pairs is due to the
  inclusion of close and bright binary satellite pairs, as suggested
  by \ibataThree{}. The \lcdm{} data shows that the same excess of
  correlated pairs independently of the brightness of the pair
  members. The excess is roughly the same even when considering only
  pairs in which both members have a magnitude difference with respect
  to the primary, $\Delta M_\rmn{r}^\rmn{Sat-Cen}$, in the range
  $1\le\Delta M_\rmn{r}^\rmn{Sat-Cen}\le 2$. Thus, contrary to the
  assertion by \ibataThree{}, close and bright binary satellites do
  not reduce the signature of a rotating disc of satellites. Discarding pairs in which one member has $\Delta
  M_\rmn{r}^\rmn{Sat-Cen}\le 2$, as suggested by \ibataThree{}, serves
  only to lower the statistics without gaining any useful information.}

\vspace{0.5cm} \MCn{In conclusion, we find that the proposals of
  \ibataThree{} cannot explain why the significance of the result
  obtained by \ibata{} decreases when varying the sample selection
  criteria. In particular, we have explicitly shown that the absence
  of an excess of anticorrelated over correlated velocity pairs for
  $V_\rmn{max}>300\kms$ and $R_\rmn{max}>150\kpc$ cannot be due to an
  increase in the contamination rate by interloper galaxies, as
  claimed by \ibataThree{}.  Similarly, we have shown that the absence
  of a rotating disc signature for same-side pairs cannot be due to
  close and bright binary satellites since such pairs, in fact,
  enhance, not reduce, the signature of rotating satellite
  systems. Furthermore, we find that the high spatial anisotropy in
  the spatial distribution of satellites reported by \ibataThree{}
  exceeds the expectation of the most extreme case, when all
  satellites are distributed in an infinitely thin plane.  This
  unphysical result is most likely due to an overestimation of the
  background contamination fraction.}


\section{Discussion and conclusions}
\label{sec:conclusion}
In the first part of this study we characterised the spatial
distribution of satellites in a large sample of SDSS galaxies. Our
analysis focused on isolated primaries that have one or more
satellites with spectroscopic redshifts. We used the photometric
catalogue of SDSS/DR8 galaxies to count the number of satellites as a
function of the angle they subtend relative to a reference axis
defined by the brightest satellite. We considered three samples of
primary galaxies centred on absolute magnitudes of $M_\rmn{r}=-23,$
-22 and -21. We found a clear signal of anisotropy in the spatial
distributions of satellites of the two brightest samples of primaries,
while for the faintest sample the uncertainties are of the same order
as the expected signal. We compared the observational data to the
predictions of the semi-analytic galaxy formation model of
\citet{Guo2011_SAM} implemented in the \lcdm{} Millennium and
Millennium-II cosmological simulations, and find very good agreement
between the observations and the theoretical predictions.

In the second part of this study we extended the analysis of \ibata{}
to explore if the anisotropy we detected could be related to the
rotating discs of satellites claimed by these authors.  We concluded
that the observational sample is not robust enough to detect such
discs. 
\MCnn{However, we stress that rotating satellite systems do exist in \lcdm{}}
\citep{Lovell2011} but not to the extent
reported by \ibata{}. In particular, we found the excess of
diametrically opposed pairs with anticorrelated velocities
seen by \ibata{} to be very sensitive to the sample selection
criteria. Small variations from the reference criteria employed by
these authors lead to smaller excesses of anticorrelated pairs and,
almost invariably, to a reduced significance, which in many
cases is well below $3\sigma$.  

\MCn{In general, when the selection criteria applied by \ibata{} are
  relaxed, for example by extending the radial acceptance range, the
  additional pairs found show no signal of rotation at all. Thus,
  increasing the maximum radial extent, $R_\rmn{max}$, from the
  reference value of $150$ to $300\kpc$ adds 57 new satellite pairs
  (compared to 23 in the reference sample), of which 28 have
  anticorrelated velocities and 29 have correlated velocities.
  Similarly, increasing the maximum line-of-sight velocity difference
  relative to the host, $V_\rmn{max}$, from the reference value of
  $300$ to $500\kms$ adds 29 pairs, of which 14 have anticorrelated
  velocities and 15 correlated velocities. }

\begin{figure}
     \centering
     \includegraphics[width=1.05\linewidth,angle=0]{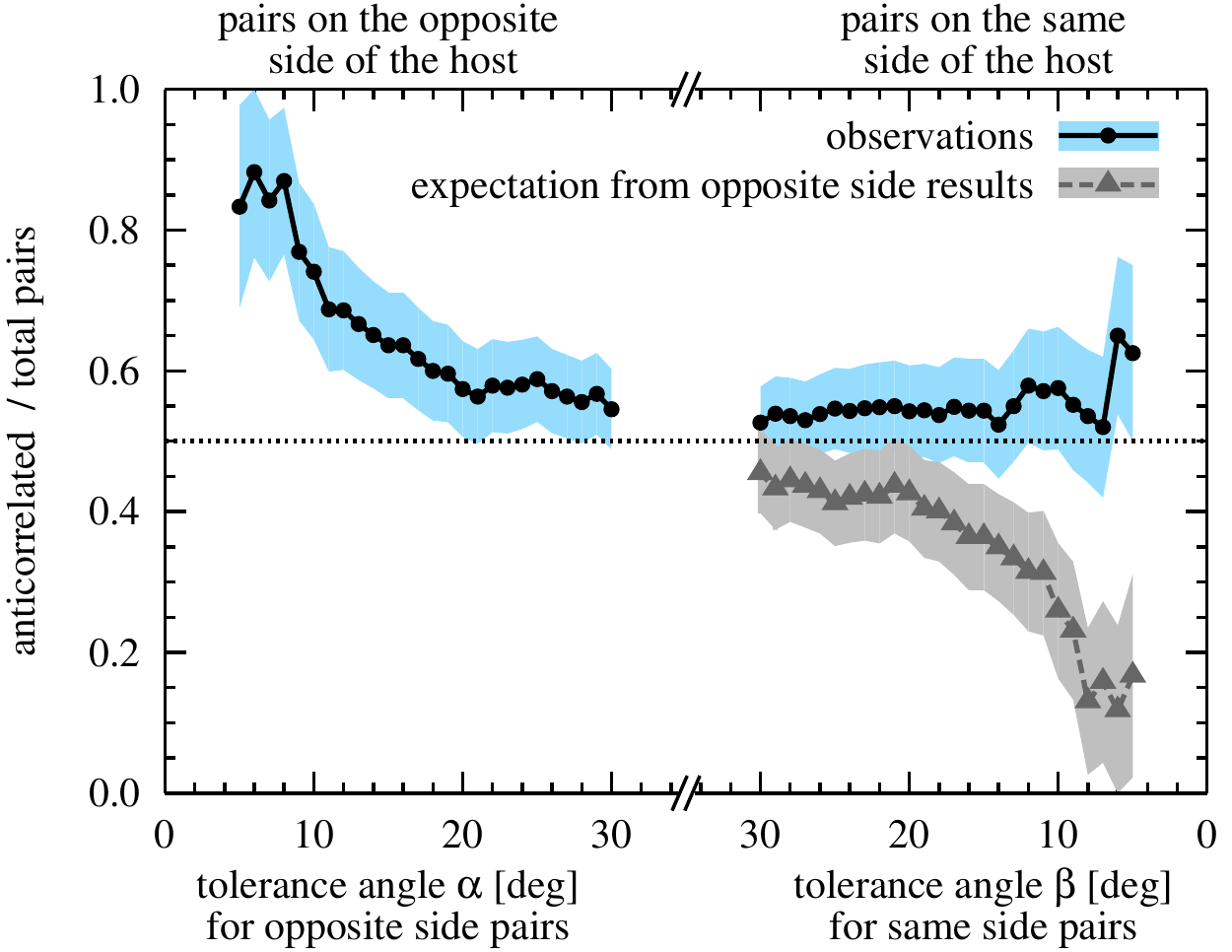}
     \caption{ Comparison of the excess of satellite pairs with
       anticorrelated velocities for pairs diametrically opposite
       (left-half) and on the same side (right-half) of the
       primary. The observational results are shown by the filled
       circles. The filled triangles show the expected signal if the
       results found for diametrically opposite pairs were indicative
       of a rotating disc of satellites (it is a mirror image of the
       left-half results with respect to the $y=x$ diagonal). The
       shaded region shows the $1\sigma$ uncertainty.  }
     \label{fig:discs_overview}
\end{figure}  

  \MCn{Thus, the reason why perturbed samples still appear to show a
  significant signal, albeit not as significant as the reference
  sample, is simply that all these samples are all correlated and
  include most, if not all, of the 23 pairs in the reference sample
  responsible for the original excess of anticorrelated over
  correlated pairs. The absence of any signal outside the reference
  sample cannot be attributed to interloper contamination, as we
  showed in \refsec{sec:ibata}.  The sensitivity of the rotation
  signal to the sample selection criteria leads us to conclude that
  the claimed detection of rotating satellite discs in the SDSS data
  is not robust.}

  To test further if the reported excess of anticorrelated velocities
  among satellites on opposite sides of the primary could originate
  from a large fraction of systems having rotating discs of
  satellites, we compared it to the expected excess of correlated
  velocities among satellites on the same side of their respective
  primaries. Using similar selection criteria to those used by
  \ibata{} to define the opposite side pairs, we found no excess of
  correlated velocities in same side pairs in the SDSS sample.
  \MCn{The absence of such an excess cannot be attributed to confusion
    introduced by the inclusion of bright binary satellites, as we
  showed in \refsec{sec:ibata}.}

The results for opposite and same side satellite systems are
summarised in \reffig{fig:discs_overview}, which shows the fraction of
anticorrelated pairs of satellites on either the opposite or the same
side of their host primary. Filled circles show the actual
measurements on both sides, while on the right half, grey triangles
denote the expected signal for same side pairs if the excess of
anticorrelated velocities measured for diametrically opposed pairs were
indicative of rotating discs. The measurements for same side pairs are
clearly in disagreement with this hypothesis, especially at small
tolerance angle, $\beta$, where the signal of a rotating disc is
expected to be maximal. This lack of any rotation signal among
same side satellites and the discrepancy with the reported signal from
opposite side satellites further weakens the evidence for universally
rotating satellite systems.

\MCn{While exposing the lack of robustness of the detection claimed by
  \ibata{}, our analysis cannot exclude the possibility that the
  ${\sim}3.5\sigma$ excess of anticorrelated over correlated pairs
  that they found is indeed a signature of rotating satellite
  systems. Such systems would have to have a projected radial extent
  of $150\kpc$ and a maximum line-of-sight velocity difference
  relative to the host of $300\kms$, since the signal is much reduced
  when either of these parameters is varied. We find this possibility
  rather unlikely. The choice of these parameters by \ibata{} was
  motivated by reference to earlier work on the PAnDAS survey of M31
  whose footprint extends to $150\kpc$, while the velocity threshold,
  $V_\rmn{max}=300\kms$, corresponds to twice the central velocity
  dispersion of M31. Not only are these choices arbitrary, but their
  relevance is unclear given that the primaries in the sample of
  \ibata{} are all much brighter than M31. We are led to conclude that
  detection by \ibata{} represents a ${\sim}3.5\sigma$
  statistical fluctuation, a conclusion that is further strengthen by
  the complete absence of a signal for same-side pairs even in their
  own sample.}

The spatial and kinematic distributions of the satellites around the
Milky Way and, more recently, around Andromeda \MCn{have been deemed a
 a serious challenge to the \lcdm{} model by several recent authors} 
\citep[e.g.][]{Kroupa2005,Pawlowski2012,Ibata2013} on the grounds that
\lcdm{} haloes seldom have satellite distributions that are as
flattened and showing the same degree of coherent rotation as found in
the Local Group
\citep{Wang2013,Bahl2014,Ibata2014b}. Our own analysis of the spatial
and kinematic distributions of the satellites around a large sample of
SDSS galaxies, returns results that are \MCn{generally} in very good
agreement with \lcdm{} predictions. 
\MCn{The satellite systems are indeed
flattened and exhibit a moderate degree of coherent
rotation. According to \lcdm{} simulations, these properties reflect
the accretion of satellites along filaments of the cosmic web. Further
characterization of satellites systems and of the cosmic web, together
with increasingly realistic cosmological simulations, should reveal
the nature of this connection in greater detail.}

\section*{Acknowledgements}

We are grateful to Rodrigo Ibata for useful and helpful discussions;
and to Shaun Cole, Wojtek Hellwing and Qi Guo for discussions during
the early stages of the project.  \MCn{We thank the anonymous referee
  for a careful reading of our paper and for detailed comments that
  have helped us improve it.}  This work was supported in part by ERC
Advanced Investigator grant COSMIWAY [grant number GA 267291] and the
Science and Technology Facilities Council [grant number ST/F001166/1,
ST/I00162X/1]. WW is supported by JRF grant number RF040353.  This
work used the DiRAC Data Centric system at Durham University, operated
by ICC on behalf of the STFC DiRAC HPC Facility
(www.dirac.ac.uk). This equipment was funded by BIS National
E-infrastructure capital grant ST/K00042X/1, STFC capital grant
ST/H008519/1, and STFC DiRAC Operations grant ST/K003267/1 and Durham
University. DiRAC is part of the National E-Infrastructure.  This
research was carried out with the support of the ``HPC Infrastructure
for Grand Challenges of Science and Engineering'' Project, co-financed
by the European Regional Development Fund under the Innovative Economy
Operational Programme.

\hypertarget{labelHypertarget}{}

\newcommand{\jcap}{JCAP} 
\newcommand{\pasa}{PASA}
\bibliographystyle{mn2e}
\bibliography{plane_reference}

\begin{thebibliography}{52}
\expandafter\ifx\csname natexlab\endcsname\relax\def\natexlab#1{#1}\fi

\bibitem[{{Abazajian} {et~al}\mbox{.}(2009){Abazajian}, {Adelman-McCarthy},
  {Ag{\"u}eros}, {Allam}, {Allende Prieto}, {An}, {Anderson}, {Anderson},
  {Annis}, {Bahcall}, \& et~al.}]{Abazajian2009}
{Abazajian} K.~N. {et~al.}, 2009, \apjs, 182, 543

\bibitem[{{Agustsson} \& {Brainerd}(2010)}]{Agustsson2010}
{Agustsson} I., {Brainerd} T.~G., 2010, \apj, 709, 1321

\bibitem[{{Aihara} {et~al}\mbox{.}(2011){Aihara}, {Allende Prieto}, {An},
  {Anderson}, {Aubourg}, {Balbinot}, {Beers}, {Berlind}, {Bickerton},
  {Bizyaev}, {Blanton}, {Bochanski}, {Bolton}, {Bovy}, {Brandt}, {Brinkmann},
  {Brown}, {Brownstein}, {Busca}, {Campbell}, {Carr}, {Chen}, {Chiappini},
  {Comparat}, {Connolly}, {Cortes}, {Croft}, {Cuesta}, {da Costa}, {Davenport},
  {Dawson}, {Dhital}, {Ealet}, {Ebelke}, {Edmondson}, {Eisenstein},
  {Escoffier}, {Esposito}, {Evans}, {Fan}, {Femen{\'{\i}}a Castell{\'a}},
  {Font-Ribera}, {Frinchaboy}, {Ge}, {Gillespie}, {Gilmore}, {Gonz{\'a}lez
  Hern{\'a}ndez}, {Gott}, {Gould}, {Grebel}, {Gunn}, {Hamilton}, {Harding},
  {Harris}, {Hawley}, {Hearty}, {Ho}, {Hogg}, {Holtzman}, {Honscheid}, {Inada},
  {Ivans}, {Jiang}, {Johnson}, {Jordan}, {Jordan}, {Kazin}, {Kirkby}, {Klaene},
  {Knapp}, {Kneib}, {Kochanek}, {Koesterke}, {Kollmeier}, {Kron}, {Lampeitl},
  {Lang}, {Le Goff}, {Lee}, {Lin}, {Long}, {Loomis}, {Lucatello}, {Lundgren},
  {Lupton}, {Ma}, {MacDonald}, {Mahadevan}, {Maia}, {Makler}, {Malanushenko},
  {Malanushenko}, {Mandelbaum}, {Maraston}, {Margala}, {Masters}, {McBride},
  {McGehee}, {McGreer}, {M{\'e}nard}, {Miralda-Escud{\'e}}, {Morrison},
  {Mullally}, {Muna}, {Munn}, {Murayama}, {Myers}, {Naugle}, {Neto}, {Nguyen},
  {Nichol}, {O'Connell}, {Ogando}, {Olmstead}, {Oravetz}, {Padmanabhan},
  {Palanque-Delabrouille}, {Pan}, {Pandey}, {P{\^a}ris}, {Percival},
  {Petitjean}, {Pfaffenberger}, {Pforr}, {Phleps}, {Pichon}, {Pieri}, {Prada},
  {Price-Whelan}, {Raddick}, {Ramos}, {Reyl{\'e}}, {Rich}, {Richards}, {Rix},
  {Robin}, {Rocha-Pinto}, {Rockosi}, {Roe}, {Rollinde}, {Ross}, {Ross},
  {Rossetto}, {S{\'a}nchez}, {Sayres}, {Schlegel}, {Schlesinger}, {Schmidt},
  {Schneider}, {Sheldon}, {Shu}, {Simmerer}, {Simmons}, {Sivarani}, {Snedden},
  {Sobeck}, {Steinmetz}, {Strauss}, {Szalay}, {Tanaka}, {Thakar}, {Thomas},
  {Tinker}, {Tofflemire}, {Tojeiro}, {Tremonti}, {Vandenberg}, {Vargas
  Maga{\~n}a}, {Verde}, {Vogt}, {Wake}, {Wang}, {Weaver}, {Weinberg}, {White},
  {White}, {Yanny}, {Yasuda}, {Yeche}, \& {Zehavi}}]{Aihara2011}
{Aihara} H. {et~al.}, 2011, \apjs, 193, 29

\bibitem[{{Bahl} \& {Baumgardt}(2014)}]{Bahl2014}
{Bahl} H., {Baumgardt} H., 2014, \mnras, 438, 2916

\bibitem[{{Bailin} {et~al}\mbox{.}(2008){Bailin}, {Power}, {Norberg},
  {Zaritsky}, \& {Gibson}}]{Bailin2008}
{Bailin} J., {Power} C., {Norberg} P., {Zaritsky} D., {Gibson} B.~K., 2008,
  \mnras, 390, 1133

\bibitem[{{Blanton} {et~al}\mbox{.}(2005){Blanton}, {Schlegel}, {Strauss},
  {Brinkmann}, {Finkbeiner}, {Fukugita}, {Gunn}, {Hogg}, {Ivezi{\'c}}, {Knapp},
  {Lupton}, {Munn}, {Schneider}, {Tegmark}, \& {Zehavi}}]{Blanton2005}
{Blanton} M.~R. {et~al.}, 2005, \aj, 129, 2562

\bibitem[{{Boylan-Kolchin} {et~al}\mbox{.}(2009){Boylan-Kolchin}, {Springel},
  {White}, {Jenkins}, \& {Lemson}}]{Boylan-Kolchin2009}
{Boylan-Kolchin} M., {Springel} V., {White} S.~D.~M., {Jenkins} A., {Lemson}
  G., 2009, \mnras, 398, 1150

\bibitem[{{Brainerd}(2005)}]{Brainerd2005}
{Brainerd} T.~G., 2005, \apjl, 628, L101

\bibitem[{{Crain} {et~al}\mbox{.}(2009){Crain}, {Theuns}, {Dalla Vecchia},
  {Eke}, {Frenk}, {Jenkins}, {Kay}, {Peacock}, {Pearce}, {Schaye}, {Springel},
  {Thomas}, {White}, \& {Wiersma}}]{Crain2009}
{Crain} R.~A. {et~al.}, 2009, \mnras, 399, 1773

\bibitem[{{Cunha} {et~al}\mbox{.}(2009){Cunha}, {Lima}, {Oyaizu}, {Frieman}, \&
  {Lin}}]{Cunha2009}
{Cunha} C.~E., {Lima} M., {Oyaizu} H., {Frieman} J., {Lin} H., 2009, \mnras,
  396, 2379

\bibitem[{{Deason} {et~al}\mbox{.}(2011){Deason}, {McCarthy}, {Font}, {Evans},
  {Frenk}, {Belokurov}, {Libeskind}, {Crain}, \& {Theuns}}]{Deason2011}
{Deason} A.~J. {et~al.}, 2011, \mnras, 415, 2607

\bibitem[{{Forero-Romero} {et~al}\mbox{.}(2011){Forero-Romero}, {Hoffman},
  {Yepes}, {Gottl{\"o}ber}, {Piontek}, {Klypin}, \&
  {Steinmetz}}]{Forero-Romero2011}
{Forero-Romero} J.~E., {Hoffman} Y., {Yepes} G., {Gottl{\"o}ber} S., {Piontek}
  R., {Klypin} A., {Steinmetz} M., 2011, \mnras, 417, 1434

\bibitem[{{Fouquet} {et~al}\mbox{.}(2012){Fouquet}, {Hammer}, {Yang}, {Puech},
  \& {Flores}}]{Fouquet2012}
{Fouquet} S., {Hammer} F., {Yang} Y., {Puech} M., {Flores} H., 2012, \mnras,
  427, 1769

\bibitem[{{Guo} {et~al}\mbox{.}(2012){Guo}, {Cole}, {Eke}, \&
  {Frenk}}]{Guo2012}
{Guo} Q., {Cole} S., {Eke} V., {Frenk} C., 2012, \mnras, 427, 428

\bibitem[{{Guo} {et~al}\mbox{.}(2011){Guo}, {White}, {Boylan-Kolchin}, {De
  Lucia}, {Kauffmann}, {Lemson}, {Li}, {Springel}, \& {Weinmann}}]{Guo2011_SAM}
{Guo} Q. {et~al.}, 2011, \mnras, 413, 101

\bibitem[{{Hammer} {et~al}\mbox{.}(2013){Hammer}, {Yang}, {Fouquet},
  {Pawlowski}, {Kroupa}, {Puech}, {Flores}, \& {Wang}}]{Hammer2013}
{Hammer} F., {Yang} Y., {Fouquet} S., {Pawlowski} M.~S., {Kroupa} P., {Puech}
  M., {Flores} H., {Wang} J., 2013, \mnras, 431, 3543

\bibitem[{{Ibata} {et~al}\mbox{.}(2014{\natexlab{a}}){Ibata}, {Ibata},
  {Famaey}, \& {Lewis}}]{Ibata2014}
{Ibata} N.~G., {Ibata} R.~A., {Famaey} B., {Lewis} G.~F., 2014{\natexlab{a}},
  \nat, 511, 563, (\ibataTwo{})

\bibitem[{{Ibata} {et~al}\mbox{.}(2014{\natexlab{b}}){Ibata}, {Famaey},
  {Lewis}, {Ibata}, \& {Martin}}]{Ibata2014d}
{Ibata} R.~A., {Famaey} B., {Lewis} G.~F., {Ibata} N.~G., {Martin} N.,
  2014{\natexlab{b}}, preprints arXiv:1411.3718, (\ibataFourth{})

\bibitem[{{Ibata} {et~al}\mbox{.}(2014{\natexlab{c}}){Ibata}, {Ibata}, {Lewis},
  {Martin}, {Conn}, {Elahi}, {Arias}, \& {Fernando}}]{Ibata2014b}
{Ibata} R.~A., {Ibata} N.~G., {Lewis} G.~F., {Martin} N.~F., {Conn} A., {Elahi}
  P., {Arias} V., {Fernando} N., 2014{\natexlab{c}}, \apjl, 784, L6

\bibitem[{{Ibata} {et~al}\mbox{.}(2013){Ibata}, {Lewis}, {Conn}, {Irwin},
  {McConnachie}, {Chapman}, {Collins}, {Fardal}, {Ferguson}, {Ibata}, {Mackey},
  {Martin}, {Navarro}, {Rich}, {Valls-Gabaud}, \& {Widrow}}]{Ibata2013}
{Ibata} R.~A. {et~al.}, 2013, \nat, 493, 62

\bibitem[{{Koch} \& {Grebel}(2006)}]{Koch2006}
{Koch} A., {Grebel} E.~K., 2006, \aj, 131, 1405

\bibitem[{{Kroupa}(2012)}]{Kroupa2012}
{Kroupa} P., 2012, \pasa, 29, 395

\bibitem[{{Kroupa}, {Theis} \& {Boily}(2005){Kroupa}, {Theis}, \&
  {Boily}}]{Kroupa2005}
{Kroupa} P., {Theis} C., {Boily} C.~M., 2005, \aap, 431, 517

\bibitem[{{Lares}, {Lambas} \& {Dom{\'{\i}}nguez}(2011){Lares}, {Lambas}, \&
  {Dom{\'{\i}}nguez}}]{Lares2011}
{Lares} M., {Lambas} D.~G., {Dom{\'{\i}}nguez} M.~J., 2011, \aj, 142, 13

\bibitem[{{Li} \& {Helmi}(2008)}]{Li2008}
{Li} Y.-S., {Helmi} A., 2008, \mnras, 385, 1365

\bibitem[{{Libeskind} {et~al}\mbox{.}(2005){Libeskind}, {Frenk}, {Cole},
  {Helly}, {Jenkins}, {Navarro}, \& {Power}}]{Libeskind2005}
{Libeskind} N.~I., {Frenk} C.~S., {Cole} S., {Helly} J.~C., {Jenkins} A.,
  {Navarro} J.~F., {Power} C., 2005, \mnras, 363, 146

\bibitem[{{Libeskind} {et~al}\mbox{.}(2009){Libeskind}, {Frenk}, {Cole},
  {Jenkins}, \& {Helly}}]{Libeskind2009}
{Libeskind} N.~I., {Frenk} C.~S., {Cole} S., {Jenkins} A., {Helly} J.~C., 2009,
  \mnras, 399, 550

\bibitem[{{Libeskind} {et~al}\mbox{.}(2014){Libeskind}, {Knebe}, {Hoffman}, \&
  {Gottl{\"o}ber}}]{Libeskind2014}
{Libeskind} N.~I., {Knebe} A., {Hoffman} Y., {Gottl{\"o}ber} S., 2014, \mnras,
  443, 1274

\bibitem[{{Libeskind} {et~al}\mbox{.}(2011){Libeskind}, {Knebe}, {Hoffman},
  {Gottl{\"o}ber}, {Yepes}, \& {Steinmetz}}]{Libeskind2011}
{Libeskind} N.~I., {Knebe} A., {Hoffman} Y., {Gottl{\"o}ber} S., {Yepes} G.,
  {Steinmetz} M., 2011, \mnras, 411, 1525

\bibitem[{{Lovell} {et~al}\mbox{.}(2011){Lovell}, {Eke}, {Frenk}, \&
  {Jenkins}}]{Lovell2011}
{Lovell} M.~R., {Eke} V.~R., {Frenk} C.~S., {Jenkins} A., 2011, \mnras, 413,
  3013

\bibitem[{{Lynden-Bell}(1976)}]{Lynden-Bell1976}
{Lynden-Bell} D., 1976, \mnras, 174, 695

\bibitem[{{McConnachie} \& {Irwin}(2006)}]{McConnachie2006}
{McConnachie} A.~W., {Irwin} M.~J., 2006, \mnras, 365, 902

\bibitem[{{McConnachie} {et~al}\mbox{.}(2009){McConnachie}, {Irwin}, {Ibata},
  {Dubinski}, {Widrow}, {Martin}, {C{\^o}t{\'e}}, {Dotter}, {Navarro},
  {Ferguson}, {Puzia}, {Lewis}, {Babul}, {Barmby}, {Bienaym{\'e}}, {Chapman},
  {Cockcroft}, {Collins}, {Fardal}, {Harris}, {Huxor}, {Mackey},
  {Pe{\~n}arrubia}, {Rich}, {Richer}, {Siebert}, {Tanvir}, {Valls-Gabaud}, \&
  {Venn}}]{McConnachie2009}
{McConnachie} A.~W. {et~al.}, 2009, \nat, 461, 66

\bibitem[{{Metz}, {Kroupa} \& {Jerjen}(2009){Metz}, {Kroupa}, \&
  {Jerjen}}]{Metz2009}
{Metz} M., {Kroupa} P., {Jerjen} H., 2009, \mnras, 394, 2223

\bibitem[{{Metz}, {Kroupa} \& {Libeskind}(2008){Metz}, {Kroupa}, \&
  {Libeskind}}]{Metz2008c}
{Metz} M., {Kroupa} P., {Libeskind} N.~I., 2008, \apj, 680, 287

\bibitem[{{Metz} {et~al}\mbox{.}(2009){Metz}, {Kroupa}, {Theis}, {Hensler}, \&
  {Jerjen}}]{Metz2009b}
{Metz} M., {Kroupa} P., {Theis} C., {Hensler} G., {Jerjen} H., 2009, \apj, 697,
  269

\bibitem[{{Nierenberg} {et~al}\mbox{.}(2012){Nierenberg}, {Auger}, {Treu},
  {Marshall}, {Fassnacht}, \& {Busha}}]{Nierenberg2012}
{Nierenberg} A.~M., {Auger} M.~W., {Treu} T., {Marshall} P.~J., {Fassnacht}
  C.~D., {Busha} M.~T., 2012, \apj, 752, 99

\bibitem[{{Pawlowski} {et~al}\mbox{.}(2014){Pawlowski}, {Famaey}, {Jerjen},
  {Merritt}, {Kroupa}, {Dabringhausen}, {L{\"u}ghausen}, {Forbes}, {Hensler},
  {Hammer}, {Puech}, {Fouquet}, {Flores}, \& {Yang}}]{Pawlowski2014c}
{Pawlowski} M.~S. {et~al.}, 2014, \mnras, 442, 2362

\bibitem[{{Pawlowski} \& {Kroupa}(2013)}]{Pawlowski2013b}
{Pawlowski} M.~S., {Kroupa} P., 2013, \mnras, 435, 2116

\bibitem[{{Pawlowski} {et~al}\mbox{.}(2012){Pawlowski}, {Kroupa}, {Angus}, {de
  Boer}, {Famaey}, \& {Hensler}}]{Pawlowski2012b}
{Pawlowski} M.~S., {Kroupa} P., {Angus} G., {de Boer} K.~S., {Famaey} B.,
  {Hensler} G., 2012, \mnras, 424, 80

\bibitem[{{Pawlowski} \& {McGaugh}(2014)}]{Pawlowski2014d}
{Pawlowski} M.~S., {McGaugh} S.~S., 2014, \apjl, 789, L24

\bibitem[{{Pawlowski}, {Pflamm-Altenburg} \& {Kroupa}(2012){Pawlowski},
  {Pflamm-Altenburg}, \& {Kroupa}}]{Pawlowski2012}
{Pawlowski} M.~S., {Pflamm-Altenburg} J., {Kroupa} P., 2012, \mnras, 423, 1109

\bibitem[{{Shaw} {et~al}\mbox{.}(2006){Shaw}, {Weller}, {Ostriker}, \&
  {Bode}}]{Shaw2006}
{Shaw} L.~D., {Weller} J., {Ostriker} J.~P., {Bode} P., 2006, \apj, 646, 815

\bibitem[{{Springel} {et~al}\mbox{.}(2008){Springel}, {Wang}, {Vogelsberger},
  {Ludlow}, {Jenkins}, {Helmi}, {Navarro}, {Frenk}, \& {White}}]{Springel2008}
{Springel} V. {et~al.}, 2008, \mnras, 391, 1685

\bibitem[{{Springel} {et~al}\mbox{.}(2005){Springel}, {White}, {Jenkins},
  {Frenk}, {Yoshida}, {Gao}, {Navarro}, {Thacker}, {Croton}, {Helly},
  {Peacock}, {Cole}, {Thomas}, {Couchman}, {Evrard}, {Colberg}, \&
  {Pearce}}]{Springel2005}
{Springel} V. {et~al.}, 2005, \nat, 435, 629

\bibitem[{{Wang}, {Frenk} \& {Cooper}(2013){Wang}, {Frenk}, \&
  {Cooper}}]{Wang2013}
{Wang} J., {Frenk} C.~S., {Cooper} A.~P., 2013, \mnras, 429, 1502

\bibitem[{{Wang} {et~al}\mbox{.}(2014){Wang}, {Sales}, {Henriques}, \&
  {White}}]{WangW2014}
{Wang} W., {Sales} L.~V., {Henriques} B.~M.~B., {White} S.~D.~M., 2014, \mnras,
  442, 1363

\bibitem[{{Wang} \& {White}(2012)}]{WangW2012}
{Wang} W., {White} S.~D.~M., 2012, \mnras, 424, 2574, (\WangWhiteTwo{})

\bibitem[{{Warnick} \& {Knebe}(2006)}]{Warnick2006}
{Warnick} K., {Knebe} A., 2006, \mnras, 369, 1253

\bibitem[{{Yang} {et~al}\mbox{.}(2006){Yang}, {van den Bosch}, {Mo}, {Mao},
  {Kang}, {Weinmann}, {Guo}, \& {Jing}}]{Yang2006}
{Yang} X., {van den Bosch} F.~C., {Mo} H.~J., {Mao} S., {Kang} X., {Weinmann}
  S.~M., {Guo} Y., {Jing} Y.~P., 2006, \mnras, 369, 1293

\bibitem[{{Yang} {et~al}\mbox{.}(2014){Yang}, {Hammer}, {Fouquet}, {Flores},
  {Puech}, {Pawlowski}, \& {Kroupa}}]{Yang2014}
{Yang} Y., {Hammer} F., {Fouquet} S., {Flores} H., {Puech} M., {Pawlowski}
  M.~S., {Kroupa} P., 2014, \mnras, 442, 2419

\bibitem[{{Zentner} {et~al}\mbox{.}(2005){Zentner}, {Kravtsov}, {Gnedin}, \&
  {Klypin}}]{Zentner2005a}
{Zentner} A.~R., {Kravtsov} A.~V., {Gnedin} O.~Y., {Klypin} A.~A., 2005, \apj,
  629, 219

\end{thebibliography}

%
%

\end{document}